\documentclass[prb,aps,amsmath,amssymb,
twocolumn,
floatfix,
longbibliography,
superscriptaddress]{revtex4-1}
\usepackage{amsthm}
\usepackage{amsfonts}
\usepackage{siunitx}
\usepackage{amsmath}
\usepackage{amssymb}
\usepackage{graphicx}
\usepackage{verbatim}
\usepackage[colorlinks]{hyperref}
\usepackage{tikz}
\usepackage{pgfplots}
\usepackage{braket}

\definecolor{linkcolor}{RGB}{0,83,166}
\hypersetup{
  colorlinks = true,
  allcolors = {linkcolor}
}

\definecolor{lines1}{RGB}{0,114,189}
\definecolor{lines2}{RGB}{217,83,25}
\definecolor{lines3}{RGB}{237,177,32}
\definecolor{lines4}{RGB}{126,47,142}
\definecolor{lines5}{RGB}{119,172,48}
\definecolor{lines6}{RGB}{77,190,238}
\definecolor{lines7}{RGB}{162,19,47}

\pgfplotsset{compat=newest}
\usetikzlibrary{plotmarks}

\pgfplotsset{every axis/.append style={
    axis line style=thick,
  }
}

\definecolor{lines1}{RGB}{0,114,189}
\definecolor{lines2}{RGB}{217,83,25}
\definecolor{lines3}{RGB}{237,177,32}
\definecolor{lines4}{RGB}{126,47,142}
\definecolor{lines5}{RGB}{119,172,48}
\definecolor{lines6}{RGB}{77,190,238}
\definecolor{lines7}{RGB}{162,19,47}
\newlength\figurewidth
\newlength\figureheight

\begin{document}
\title{Observation of topological phenomena in a programmable lattice of 1,800 qubits} 

\author{Andrew D.~King}\email[]{aking@dwavesys.com}
\affiliation{D-Wave Systems Inc., 3033 Beta Avenue, Burnaby, BC, Canada V5G 4M9}
\author{Juan Carrasquilla}
\affiliation{Vector Institute, MaRS Centre, Toronto, ON, Canada M5G 1M1}
\author{Isil Ozfidan}
\affiliation{D-Wave Systems Inc., 3033 Beta Avenue, Burnaby, BC, Canada V5G 4M9}
\author{Jack Raymond}
\affiliation{D-Wave Systems Inc., 3033 Beta Avenue, Burnaby, BC, Canada V5G 4M9}
\author{Evgeny Andriyash}
\affiliation{D-Wave Systems Inc., 3033 Beta Avenue, Burnaby, BC, Canada V5G 4M9}
\author{Andrew Berkley}
\affiliation{D-Wave Systems Inc., 3033 Beta Avenue, Burnaby, BC, Canada V5G 4M9}
\author{Mauricio Reis}
\affiliation{D-Wave Systems Inc., 3033 Beta Avenue, Burnaby, BC, Canada V5G 4M9}
\author{Trevor M.~Lanting}
\affiliation{D-Wave Systems Inc., 3033 Beta Avenue, Burnaby, BC, Canada V5G 4M9}
\author{Richard Harris}
\affiliation{D-Wave Systems Inc., 3033 Beta Avenue, Burnaby, BC, Canada V5G 4M9}
\author{Gabriel Poulin-Lamarre}
\affiliation{D-Wave Systems Inc., 3033 Beta Avenue, Burnaby, BC, Canada V5G 4M9}
\author{Anatoly Yu.~Smirnov}
\affiliation{D-Wave Systems Inc., 3033 Beta Avenue, Burnaby, BC, Canada V5G 4M9}
\author{Christopher Rich}
\affiliation{D-Wave Systems Inc., 3033 Beta Avenue, Burnaby, BC, Canada V5G 4M9}
\author{Fabio Altomare}
\affiliation{D-Wave Systems Inc., 3033 Beta Avenue, Burnaby, BC, Canada V5G 4M9}
\author{Paul Bunyk}
\affiliation{D-Wave Systems Inc., 3033 Beta Avenue, Burnaby, BC, Canada V5G 4M9}
\author{Jed Whittaker}
\affiliation{D-Wave Systems Inc., 3033 Beta Avenue, Burnaby, BC, Canada V5G 4M9}
\author{Loren Swenson}
\affiliation{D-Wave Systems Inc., 3033 Beta Avenue, Burnaby, BC, Canada V5G 4M9}
\author{Emile Hoskinson}
\affiliation{D-Wave Systems Inc., 3033 Beta Avenue, Burnaby, BC, Canada V5G 4M9}
\author{Yuki Sato}
\affiliation{D-Wave Systems Inc., 3033 Beta Avenue, Burnaby, BC, Canada V5G 4M9}
\author{Mark Volkmann}
\affiliation{D-Wave Systems Inc., 3033 Beta Avenue, Burnaby, BC, Canada V5G 4M9}
\author{Eric Ladizinsky}
\affiliation{D-Wave Systems Inc., 3033 Beta Avenue, Burnaby, BC, Canada V5G 4M9}
\author{Mark Johnson}
\affiliation{D-Wave Systems Inc., 3033 Beta Avenue, Burnaby, BC, Canada V5G 4M9}
\author{Jeremy Hilton}
\affiliation{D-Wave Systems Inc., 3033 Beta Avenue, Burnaby, BC, Canada V5G 4M9}
\author{Mohammad H.~Amin}
\affiliation{D-Wave Systems Inc., 3033 Beta Avenue, Burnaby, BC, Canada V5G 4M9}
\affiliation{Department of Physics, Simon Fraser University, Burnaby, BC, Canada V5A 1S6, Burnaby B.C.}
\date{\today}

\maketitle

{\bf The celebrated work of Berezinskii, Kosterlitz and Thouless in the 1970s\cite{Berezinskii1972,Kosterlitz1973} revealed exotic phases of matter governed by topological properties of low-dimensional materials such as thin films of superfluids and superconductors.  Key to this phenomenon is the appearance and interaction of vortices and antivortices in an angular degree of freedom---typified by the classical XY model---due to thermal fluctuations.  In the 2D Ising model this angular degree of freedom is absent in the classical case, but with the addition of a transverse field it can emerge from the interplay between frustration and quantum fluctuations.  Consequently a Kosterlitz-Thouless (KT) phase transition has been predicted in the quantum system by theory and simulation\cite{Moessner2000,Moessner2001,Isakov2003}.  
Here we demonstrate a large-scale quantum simulation of this phenomenon in a network of 1,800 {\em in situ} programmable superconducting flux qubits arranged in a fully-frustrated square-octagonal lattice.  Essential to the critical behavior, we observe the emergence of a complex order parameter with continuous rotational symmetry, and the onset of quasi-long-range order as the system approaches a critical temperature.  We use a simple but previously undemonstrated approach to statistical estimation with an annealing-based quantum processor, performing Monte Carlo sampling in a chain of reverse quantum annealing protocols.  Observations are consistent with classical simulations across a range of Hamiltonian parameters.  We anticipate that our approach of using a quantum processor as a programmable magnetic lattice will find widespread use in the simulation and development of exotic materials.}

Richard Feynman's vision of simulating quantum systems with a quantum computer\cite{Feynman1982, Lloyd1996} has motivated the field of quantum information since its inception.  In the absence of large-scale programmable universal quantum computers, advances in quantum simulation are bound by available technology.  Still, remarkable progress has been made using near-term approaches such as ultracold Rydberg atoms\cite{Gross2017}, superconducting qubits\cite{Paraoanu2014}, trapped ions\cite{Hadzibabic2006,Zhang2017b}, and quantum dots\cite{Hensgens2017}, among others\cite{Georgescu2014}.  Due to the possibility of noise-tolerant applications, quantum simulation has been identified as a potential area of commercial value for near-term quantum computing technologies \cite{Mohseni2017,Preskill2018}.
Quantum annealing (QA) processors\cite{Johnson2011,Harris2010,Bunyk2014} can be used to simulate systems in the transverse field Ising model (TFIM) described by the Hamiltonian
\begin{equation}\label{eq:ham}
 H = \sum_{i}h_{i}\sigma_i^z + \sum_{ i <j }  J_{ij} \sigma^z_i \sigma^z_j - \Gamma\sum_{i} \sigma^x_i
\end{equation}
where $h_i$ are longitudinal fields, $J_{ij}$ are coupling terms, $\sigma_i^x$ and $\sigma_i^z$ are Pauli matrices acting on the $i$th spin, and $\Gamma$ is the transverse field. Evolution of this Hamiltonian in a low-temperature environment allows sampling of low-energy solutions, with applications to optimization and machine learning\cite{Mott2017,Amin2016,Mohseni2017}.  This in turn allows estimation of equilibrium statistics of various phases dictated by the parameters $h_i$, $J_{ij}$ and $\Gamma$ at a given temperature $T$.

Geometrically frustrated magnets\cite{Moessner2006} are systems described by spin Hamiltonians with competing terms that cannot be minimized simultaneously.  These systems give rise to a broad spectrum of exotic phases of matter.  The TFIM exhibits a particular type of topological phenomenon on certain frustrated lattices: a KT phase transition separates a disordered paramagnetic phase from a phase in which complementary topological defects---vortices and antivortices---form bound pairs, resulting in polynomial decay of correlations\cite{Berezinskii1972,Kosterlitz1973,Moessner2001}.  This phase transition has been predicted in theory and simulation, but to our knowledge never observed experimentally in the TFIM.  Other KT phase transitions have, however, been observed in many systems including superconducting films, superfluid helium films, trapped atomic gases, and hybrid tin-graphene Josephson junction arrays\cite{Hadzibabic2006,Kosterlitz2016, Han2014}.

\begin{figure}

\begin{tikzpicture}\setlength{\figurewidth}{6cm}\setlength{\figureheight}{4.5cm}\sf

\node[anchor=north] (embedded) at (0,0)
{\includegraphics[height=3.0cm]{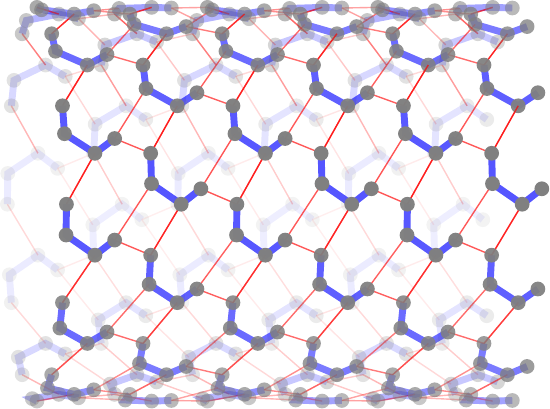}};
\node[anchor=north] (triangular) at (4,0)
{\includegraphics[height=2.9cm,width=3.3cm]{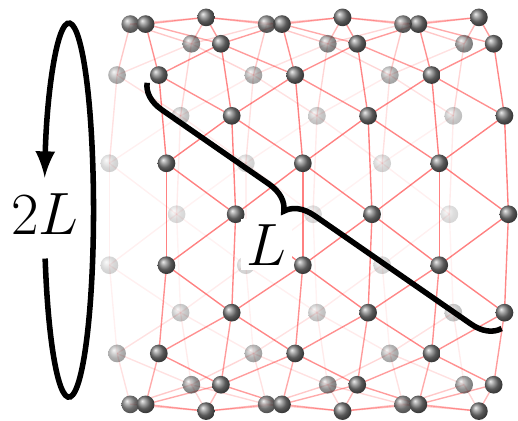}};
\node[anchor=north east] (schedule) at (4.8,5.2)
{\includegraphics[width=5.5cm]{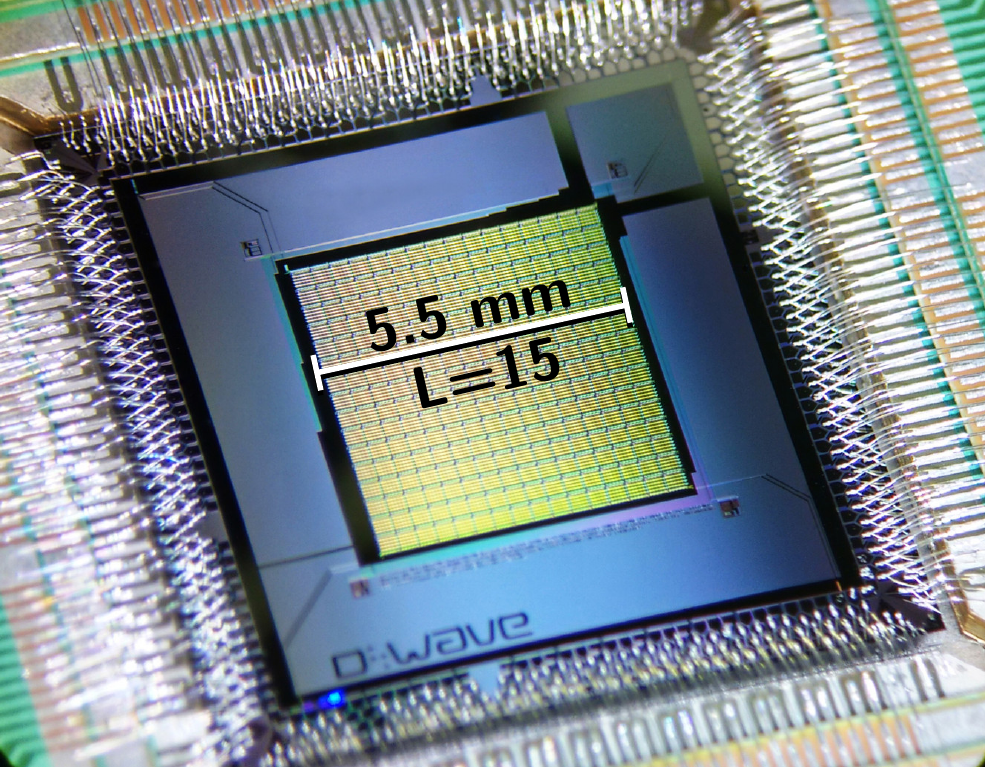}};
\node at (-2,0) {{\bfseries\sffamily\large b}};
\node at (3,0) {{\bfseries\sffamily\large c}};
\node at (-1.5,5) {{\bfseries\sffamily\large a}};
\end{tikzpicture}
\caption{{\bf Quantum annealing processor and geometrically frustrated lattices.}  {\bf a}, Photograph of a 2,048-qubit quantum annealing processor, fabricated as a superconducting integrated circuit.  {\bf b--c}, We study the fully-frustrated square-octagonal ({\bf b}) and triangular ({\bf c}) lattices with cylindrical boundary condition and width $L$ up to $15$ ($L=6$ shown).  FM couplers ($J_{ij}=-1.8$) are indicated with blue lines; AFM couplers ($J_{ij}=1$ except on the boundary, where $J_{ij}=\tfrac 12$) are indicated with red lines.
}\label{fig:lattice}
\end{figure}
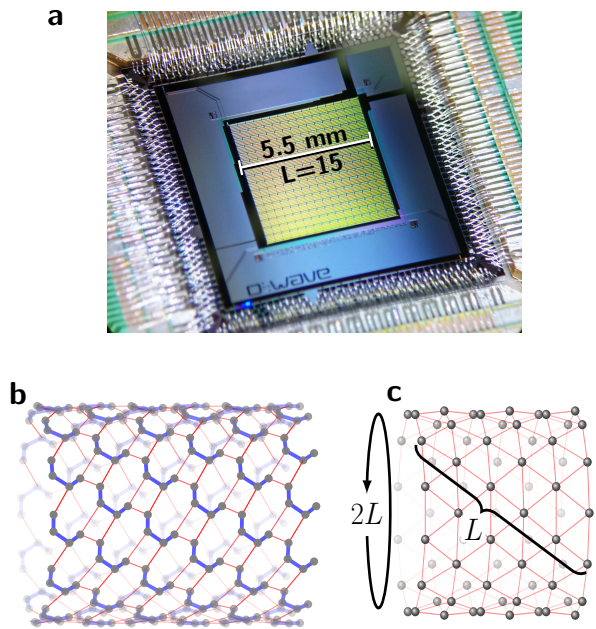

Here we exhibit this phenomenon in a large-scale programmable quantum simulation.  We perform this simulation using a superconducting QA processor consisting of 2,048 rf-SQUID flux qubits fabricated as an integrated circuit (Fig.~\ref{fig:lattice}a), in which Hamiltonian terms $h_i$ and $J_{ij}$ are specified with programmable on-chip control circuitry\cite{Johnson2011,Harris2010,Bunyk2014}.  The topology of available nonzero coupling terms $J_{ij}$ consists of a regular bipartite grid.  On this programmable substrate we define a sequence of fully-frustrated square-octagonal lattices with cylindrical boundary condition (Fig.~\ref{fig:lattice}b), with the largest system using 1,800 qubits (Supplemental \ref{sec:embedding}).  This lattice features chains of four ferromagnetically (FM) coupled qubits. In the large FM coupling limit, the low-energy description of this lattice follows the same Landau-Ginzburg-Wilson (LGW) theory as the widely studied\cite{Moessner2000,Moessner2001,Isakov2003,Jiang2005,Wang2017,Blankschtein1984a} triangular antiferromagnetic (AFM) lattice (Fig.~\ref{fig:lattice}c).  We therefore use the same analytic machinery even well outside the large FM coupling, low-temperature limit.

Central to this analysis is a mapping from each plaquette in the triangular lattice to a 2D pseudospin that can be represented by a complex number $\psi_j$.  This gives the local angular degree of freedom in which topological features, including vortices and antivortices, can emerge.  To first order perturbation in the small-$\Gamma$ limit, an AFM triangle has six degenerate ground states, each consisting of an up-spin ($\ket \uparrow$), a down-spin ($\ket\downarrow$), and a spin aligned with the transverse field ($\ket \rightarrow = (\ket \uparrow + \ket \downarrow )/\sqrt{2}$).  These six ground states map to complex six-state pseudospins via the mapping
\begin{equation}\label{eq:pseudospin}
\psi_j = \braket{\sigma_1^z} + \braket{\sigma_2^z}e^{2\pi i /3} + \braket{\sigma_3^z}e^{4\pi i /3}.
\end{equation}
As with a single AFM triangle, the triangular lattice has, to first order perturbation, a six-fold degenerate clock ground state: One third of the spins align with the transverse field, one third take spin up and one third take spin down.  This {\em order by disorder} \cite{Moessner2001}, in which the addition of quantum fluctuations brings order to a highly degenerate classical ground state manifold, naturally divides the spins into three sublattices.  Their magnetizations $m_1$, $m_2$, $m_3$, defined as the average of $\langle\sigma_i^z\rangle$ over spins in the sublattice, give a complex order parameter\cite{Blankschtein1984a}
\begin{equation}\label{eq:orderparameter}
\psi = me^{i\theta} = (m_1 + m_2e^{i2\pi/3} + m_3e^{i4\pi/3})/\sqrt{3}.
\end{equation}
In the absence of open boundary effects, $\psi$ is simply the average of pseudospin $\psi_j$ over all plaquettes in the system.  Further details are given in Supplemental \ref{sec:orderparameter}.

\begin{figure}
\begin{tikzpicture}\sf
\node[anchor=west] (vectorfield) at (0,0)
 {\includegraphics{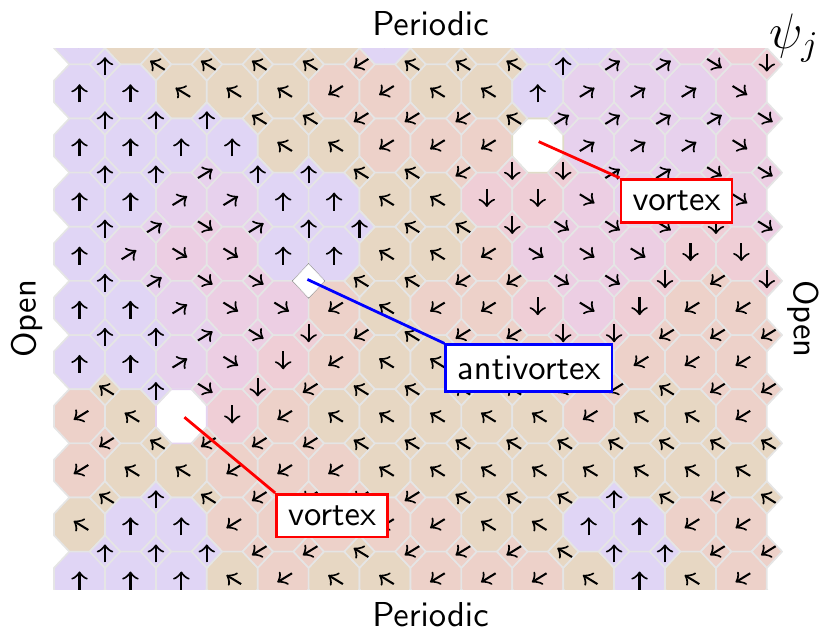}};
\node[anchor=north east] (phasediagram) at (6.9,-3.5)
{\includegraphics[scale=.82]{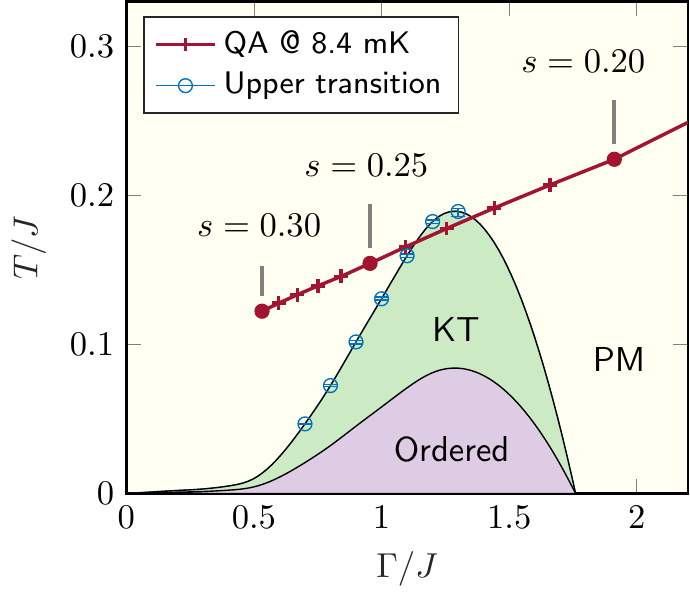}};
\node at (0.5,3) {{\bfseries\sffamily\large a}};
\node at (1,-4) {{\bfseries\sffamily\large b}};

\end{tikzpicture}\vspace{-.25cm}
\caption{{\bf Topological features in the transverse field Ising model.} {\bf a},  In the fully-frustrated square-octagonal lattice, the spins of a classical state map naturally to a complex field $\psi_j$ on the plaquettes, revealing topological features.  A swath of an output state from the QA processor, cropped at the top and bottom, is shown from this perspective, with arrows indicating values of $\psi_j$ in the complex plane.  Vortices and antivortices (marked in white) can be identified by the clockwise or anticlockwise winding of $\psi_j$ along a closed clockwise path.  These topological defects occur in the presence of all-up or all-down plaquettes and frustrated FM couplers.  Unpaired vortices and antivortices can appear due to the open boundary.  {\bf b}, The phase diagram of the square-octagonal lattice in the temperature/transverse-field ($T/\Gamma$) plane is determined by QMC simulation on toroidal systems (Supplemental \ref{sec:phasediagram}).  An ordered region and a critical KT region with $\Gamma>0$ are bounded by KT phase transitions; circles indicate one of two methods of determining the upper KT transition.  The quantum processor follows an annealing schedule that passes through the KT region at sufficiently low temperature $T \lesssim \SI{9}{mK}$.  Experiments are performed between $\SI{8.4}{mK}$ and $21.4$ mK.}\label{fig:vectorfield}
\end{figure}
In the square-octagonal lattice, we use the same order parameter with the three sublattices defined according to the natural mapping: two FM-coupled spins must be in the same sublattice, and two AFM-coupled spins must be in different sublattices.  Fig.~\ref{fig:vectorfield}a shows $\psi_j$ evaluated on each plaquette for an output state from the QA processor.

\begin{figure*}

\begin{tikzpicture}\setlength{\figurewidth}{6cm}\setlength{\figureheight}{4.5cm}\sf
\node[anchor=north] (mversuss) at (0,-0.1)
{\includegraphics[scale=.82]{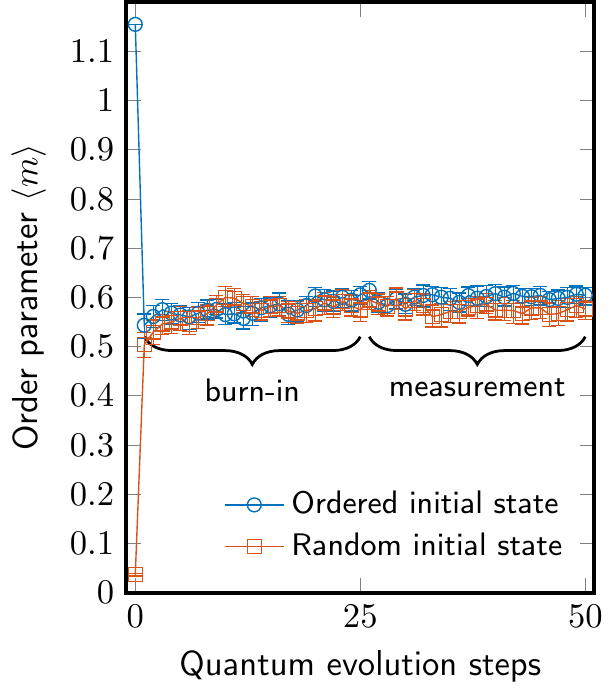}};
\node[anchor=north] (mversuss) at (0.4,-0.25)
{\includegraphics[scale=.55]{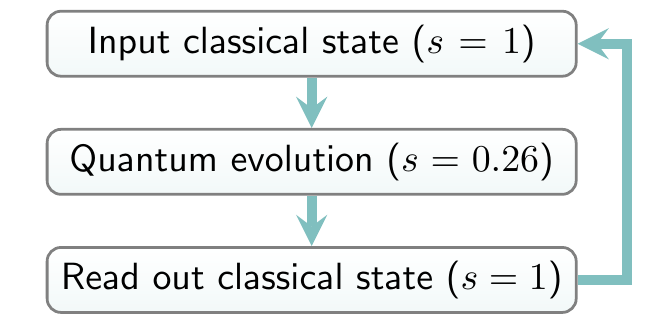}};
\node[anchor=north] (mversuss) at (5.5,-0.1)
{\includegraphics[scale=.82]{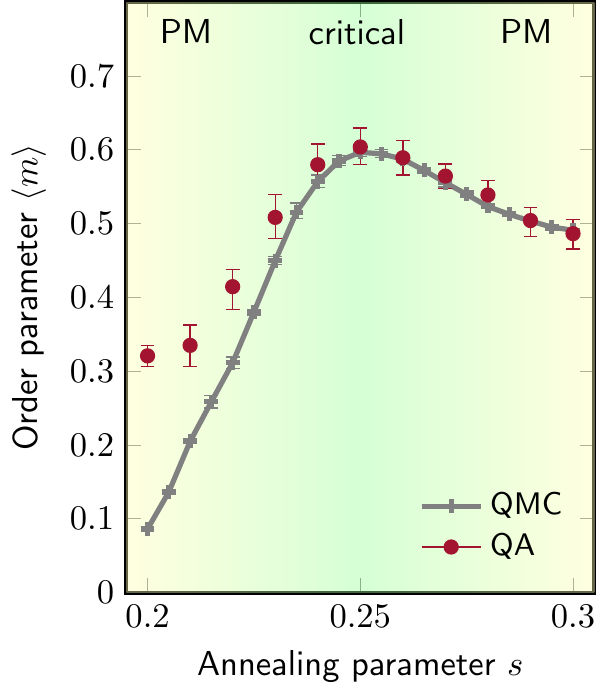}};
\node[anchor=north] (heatmaps) at (12,0)
{\includegraphics[scale=.92]{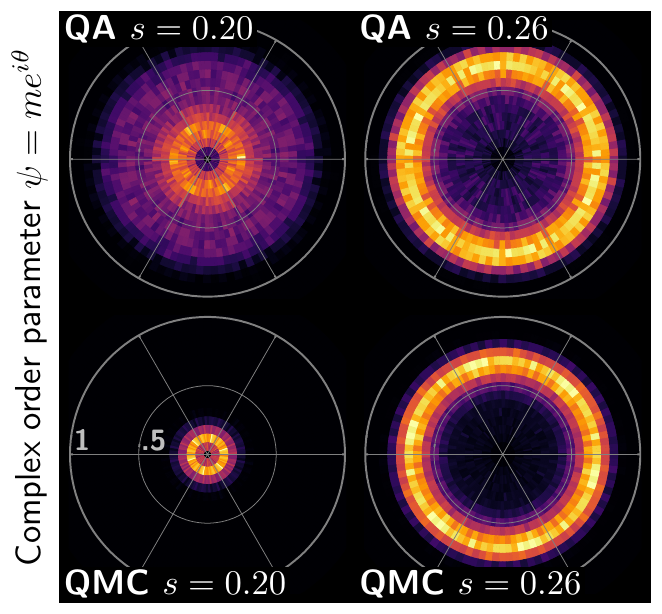}};
\node (a) at (-2,-0.2) {{\bfseries\sffamily\large a}};
\node (b) at (3.3,-0.2) {{\bfseries\sffamily\large b}};
\node (c) at (9,-0.2) {{\bfseries\sffamily\large c}};
\end{tikzpicture}\vspace{-.25cm}

\caption{{\bf Measurement of complex order parameter $\psi$ in quantum simulation.}  {\bf a}, We estimate statistics of the quantum system using a quantum evolution Monte Carlo approach: the previous classical state is input to the QA processor, evolved for $\SI{65}{ms}$ in a reverse annealing protocol (Supplemental \ref{sec:protocols}), and read out as a new classical state.  A call to the processor consists of 50 evolution steps; we discard the first 25 output samples as correlation with the initial state decays and heating from experimental setup dissipates.  Statistics (here, $\langle m\rangle = \langle|\psi|\rangle$) are estimated from the remaining 25 samples, averaged over 240 independent experiments.  Error bars are bootstrap 95\% confidence intervals.  {\bf b}, Order parameter magnitude $\langle m \rangle$ for 1,800-spin square-octagonal lattice ($L=15$).  As $s$ increases, quantum and thermal fluctuations are reduced and the system displays a peak in $\langle m \rangle$ where order is maximal.  The discrepancy between the peak location and Fig.~\ref{fig:vectorfield}b is expected from finite size effects and boundary conditions.  Estimates are close to equilibrium values as confirmed by QMC.  Overestimate of $\langle m\rangle$ from QA for small $s$, where dynamics are fast, arises from evolution during the $\SI{1}{\micro\second}$ pre-readout quench. {\bf c}, Monte Carlo histogram of the complex order parameter $\psi=me^{i\theta}$ from QA (top) and QMC (bottom) shows $U(1)$ symmetry consistent with the expected paramagnetic and KT phases ($s=0.20$ and $s=0.26$ respectively).}\label{fig:orderparameter}
\end{figure*}

Fig.~\ref{fig:vectorfield}b shows the phase diagram of the square-octagonal lattice in the $\Gamma$--$T$ plane, which resembles that of the triangular lattice\cite{Isakov2003} (Supplemental \ref{sec:phasediagram}).  At high temperature the system is in a disordered paramagnetic phase.  As $T$ drops, for $\Gamma < 1.76$, the system enters a critical phase at a KT phase transition, in which vortices and antivortices form bound pairs.  At even lower temperature the system transitions into a clock ordered phase, where $\psi$ concentrates around the six low-temperature clock states.

Having described the lattice and its phase diagram, we now turn to the task of simulating the physics with the QA processor.  The processor Hamiltonian is parameterized by a time-dependent annealing parameter $s$ that varies between $0$ and $1$, and for this particular lattice can be written as
\begin{equation}
H(s) = J(s)\sum_{i<j}J_{ij}\sigma_i^z\sigma_j^z - \Gamma(s)\sum_i\sigma_i^x
\end{equation}
where $J_{ij}$ is $1.0$ for AFM couplers and $-1.8$ for FM couplers.  Details of implementing this Hamiltonian are given in Supplemental \ref{sec:qa}.  As $s$ increases, $J(s)$ increases and $\Gamma(s)$ decreases: $J(0)\approx  0$, $J(1)\gg T$, $\Gamma(0)\gg T$, $\Gamma(1) \approx 0$.  Fig.~\ref{fig:vectorfield}b shows the set of points in the $\Gamma$--$T$ plane realized by varying $s$ at the operating temperature of $8.4 \pm\SI{0.2}{\milli\kelvin}$; points above or below this line can be reached by either rescaling $J_{ij}$ or changing the QA operating temperature.

To estimate the distribution of $\psi$ we use the QA processor as a quantum evolution update operator in a Monte Carlo method, where each evolution follows a reverse annealing protocol (Fig.~\ref{fig:orderparameter}a, details in Supplemental \ref{sec:protocols}).  This protocol involves initializing the system in a classical state at $s=1$, quickly reducing $s$ to a target value between $0.20$ and $0.30$ over approximately $\SI{3.5}{\micro s}$, and pausing for $\SI{65}{ms}$ before quenching back to $s=1$ over $\SI{1}{\micro s}$ and reading the classical output.  Repetition of this protocol resembles a Markov chain by which we generate Monte Carlo estimates of $\psi$.  This is a departure from the traditional use case of QA, in which $s$ is steadily increased from $0$ to $1$.  Typical of Markov chain Monte Carlo methods, we discard the first half of samples in each experiment as {\em burn-in} to minimize correlation with the initial state; a further motivation is that we observe around $\SI{0.1}{mK}$ of cooling at low temperatures, as heat from the QA programming cycle dissipates.  We estimate statistics using both an ordered initial state and a random intial state ($m = |\psi|$ large and small, respectively). Deviations between the two estimators are small and included in the error bars in subsequent plots.

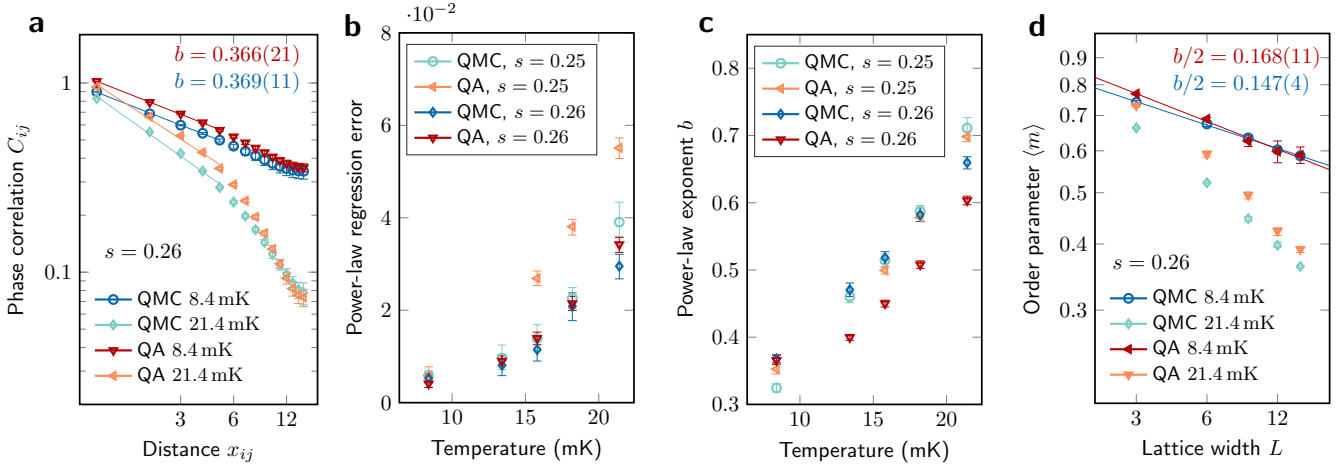
\begin{figure*}
\begin{tikzpicture}
    \node[anchor=north west] at (-.1,0)
    {\includegraphics[scale=.82]{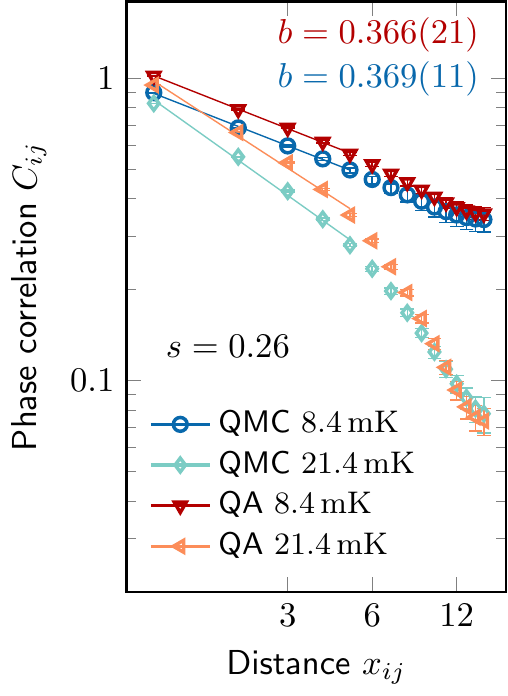}};
    \node[anchor=north west] at (4.4,.4)
    {\includegraphics[scale=.82]{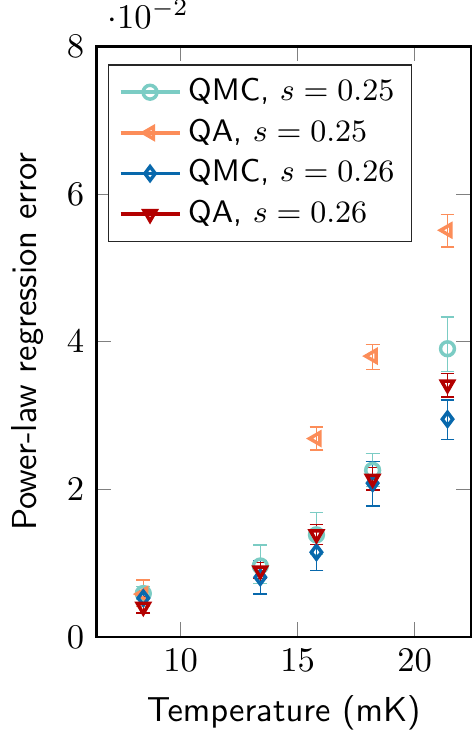}};
    \node[anchor=north west] at (8.8,0)
    {\includegraphics[scale=.82]{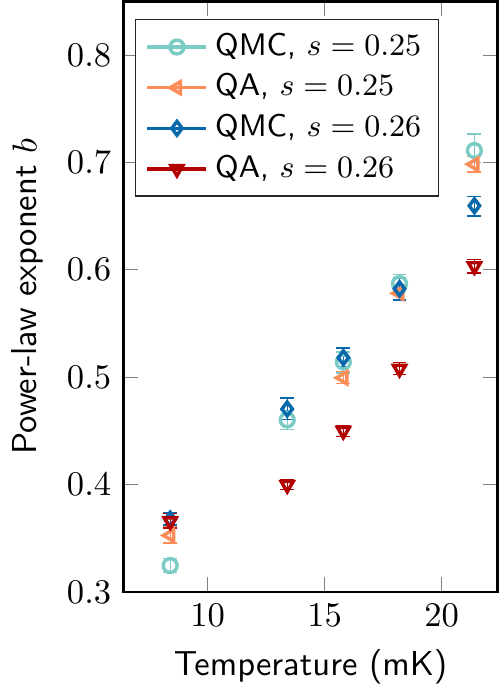}};
    \node[anchor=north west] at (13.4,0.1)
    {\includegraphics[scale=.82]{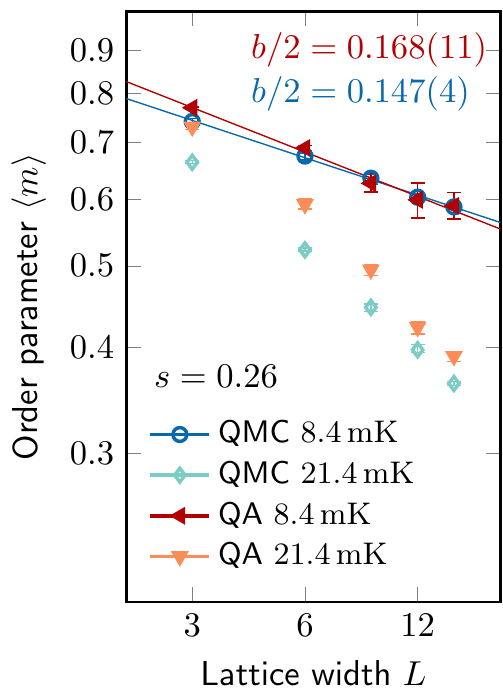}};
    \node at (0.5,0) {{\bfseries\sffamily\large a}};
    \node at (4.7,0) {{\bfseries\sffamily\large b}};
    \node at (9.4,0) {{\bfseries\sffamily\large c}};
    \node at (13.8,0) {{\bfseries\sffamily\large d}};
  \end{tikzpicture}\vspace{-.25cm}
  \caption{{\bf Power-law correlation decay near the KT crossover.}
 {\bf a}, Correlations in the complex field $\psi_j$, calculated as $C_{ij} = \langle\mathrm{Re}(\psi_i\psi_j^*)\rangle$ and measured between plaquettes along the periodic dimension far from the open boundary.  Data for QMC and QA are shown near the critical temperature ($\SI{8.4}{mK}$) and above the critical temperature ($\SI{21.4}{mK}$).  In the thermodynamic limit these correlations should decay with a power-law exponent close to $\eta=1/4$ near the critical temperature.  We perform power-law regression fit on distances $1$ to $5$, where boundary effects are minimal.  {\bf b}, As temperature is reduced, root-mean-squared regression error drops, indicating the onset of power-law decay.  {\bf c}, In both QA and QMC, the power-law exponent $b$ moves towards the expected critical value of $\eta=1/4$ as temperature decreases.  {\bf d}, Decay of $m$ versus $L$ is expected to go from exponential in the paramagnetic phase to power-law with exponent $b/2$ near the KT crossover.}\label{fig:correlation}
\end{figure*}

We experimentally probe values of $s$ between $0.20$ and $0.30$ at physical temperature $T$ between $\SI{8.4}{mK}$ and $\SI{21.4}{mK}$.  Fig.~\ref{fig:orderparameter}b shows the expectation $\langle m \rangle$ as a function of $s$ for the 1,800-spin lattice (lattice width $L=15$) at $\SI{8.4}{mK}$, compared with estimates from continuous-time path-integral quantum Monte Carlo (QMC) simulation as detailed in Supplemental \ref{sec:qmc}.  In agreement with QMC, QA shows a peak in $\langle m \rangle$ close to where the QA schedule cuts through the KT phase in the large system limit, with a small shift caused by the open boundary condition and finite size effects.  For $s \geq 0.24$, QA shows good agreement with QMC.  For smaller $s$, dynamics of the system are fast and the system orders during the $\SI{1}{\micro\second}$ quench, leading to an overestimate of $\langle m \rangle$; results are almost unchanged between $s=0$ and $s=0.2$.

The complex order parameter $\psi$ is plotted in Fig.~\ref{fig:orderparameter}c for two points along the QA schedule.  As $s$ increases from $0.20$ to $0.26$, both QA and QMC show the emergence of order: $\psi$ concentrates around a ring with the characteristic $U(1)$ symmetry of the 2D XY model despite the discrete $\mathbb Z_2$ symmetry of the Ising model\cite{Jose1977,Isakov2003,Jiang2005,Wang2017}.

In the thermodynamic limit, the KT phase transition is marked by a change of correlation decay from exponential to power-law at the critical temperature.  In the critical phase, correlations $C_{ij}$ decrease with distance $x_{ij}$ as $C_{ij} \propto x_{ij}^{-b}$ with an exponent $b$ bounded above by the universal 2D XY critical exponent  $\eta = 1/4$ \cite{Herbut2007}.  At finite sizes, the transition is expected to be broadened, and the apparent transition point can be affected by boundary conditions.  To probe the onset of critical behavior we study correlations in the complex field $\psi_j$ on the plaquettes of the largest lattice on 1,800 spins; correlation is measured as $C_{ij} = \langle\mathrm{Re}(\psi_i\psi_j^*)\rangle$ on plaquettes along the periodic dimension, halfway between the open boundaries.

Fig.~\ref{fig:correlation}a shows $C_{ij}$ as a function of distance for two temperatures at $s=0.26$, with good agreement between QMC and QA.  The absolute shift between QMC and QA correlations is explained by evolution during the quench (Supplemental \ref{sec:quench}).  As temperature decreases, correlations increase, approaching power-law decay: quality of power-law fit on distances $1$ to $5$---where boundary effects are minimal---improves as temperature decreases.  This is shown in Fig.~\ref{fig:correlation}b for $s=0.25$ and $0.26$, near the peak of $\langle m \rangle$ in $s$ (Fig.~\ref{fig:orderparameter}b).  The exponent $b$ of this power-law regression decreases with temperature (Fig.~\ref{fig:correlation}c), reaching a minimum $b \approx 0.35$ for QA and $b \approx 0.32$ for QMC.

In the vicinity of the critical region, the same onset of power-law behavior is expected for $\langle m\rangle$ as a function of size with a halved exponent; i.e., $\langle m \rangle\propto L^{-b/2}$.  This is shown in Fig.~\ref{fig:correlation}d, where QA shows agreement with QMC over a range of system sizes.  Scaling exponents of $b\approx 0.34$ and $b\approx 0.29$ for QA and QMC respectively are close to the values extracted from phase correlations in Fig.~\ref{fig:correlation}c.

We have presented a large-scale quantum simulation of an exotic phase of matter. Our experimental results show clear signatures of topological phenomena: a complex order parameter with rotational symmetry that responds as expected to changes in $\Gamma$ and $T$, and the onset of power-law scaling of correlations.  Taken together, these constitute the experimental observation of topological order in a frustrated 2D transverse field Ising model, as theoretically predicted\cite{Moessner2001,Moessner2000} and simulated in QMC\cite{Isakov2003}.
Agreement with QMC over a range of system sizes and Hamiltonian parameters---independently extracted with no fitting parameters---validates the flux qubit implementation of the transverse field Ising model at large scales.

The reverse annealing technique used in this work promises to greatly expand the utility of quantum annealing processors.  The ability for a QA processor to reverse anneal from an input state allowed us to validate convergence of our statistical estimators to a steady state, but its importance is far more general: quantum evolution of an input state is crucial to the implementation of hybrid quantum-classical algorithms\cite{chancellor2017}.

This programmable magnetic material has allowed us to study the many-body dynamics of matter that is time consuming to simulate classically and difficult to implement physically.  The methods used here can be applied to explore other exotic phases of matter in a variety of lattices.  Several future developments will improve this approach. First, greater qubit connectivity will allow more flexibility in lattice geometry and realization of fully periodic boundary conditions. Second, faster projective readout will enable more accurate sampling from systems with fast dynamics, and any advances in noise and control error will improve the accuracy of these simulations. And finally, with the addition of non-stoquastic couplings\cite{Sandvik2010}, quantum annealing processors could foreseeably simulate systems of which classical simulations are intractable even at modest scale.

\bibliography{ak_mendeley}%

\section*{Acknowledgments}

 We thank Ian Affleck, Sergio Boixo, Ed Farhi, Marcel Franz, Igor Herbut, Sergei Isakov, Seth Lloyd, Roger Melko, Masoud Mohseni, Hartmut Neven, and Yuan Wan for discussions.  We are grateful to the research, engineering, and operations staff at D-Wave Systems for directly and indirectly supporting this research.

\section*{Author contributions}

A.D.K., J.C., I.O., J.R., E.A., and M.H.A.\ conceived and designed the experiment.  J.R., E.A., I.O., and A.D.K.\ conducted classical simulations.  I.O.\ and A.D.K.\ conducted the main QA experiments.  A.B.\ and M.R.\ calibrated the QA processor.  T.M.L., R.H., G.P.L., and A.Y.S.\ conducted supporting QA expriments and theroretical analysis.  A.B., M.R., T.M.L., R.H., C.R., F.A., P.B., J.W., L.S., E.H., Y.S., M.V., E.L., M.J., and J.H.\ designed, developed, and fabricated the QA apparatus.  M.H.A., A.D.K., J.C., J.R., I.O., E.A., T.M.L., R.H., and A.Y.S.\ contributed to writing the manuscript.

\section*{Data availability}

The datasets generated and analyzed during the current study are available from the corresponding author on reasonable request.

\section*{Methods}

\subsection*{Quantum processor and experimental methods}\label{meth:qa}

The quantum annealing processor used in this work is a D-Wave 2000Q system operated in a non-annealing mode.  Operating temperatures ranging between $8.4\pm 0.2$ mK and $21.4\pm 0.2$ mK are measured with a cryostat thermometer and independently verified using single-qubit susceptibility measurements.  In the reverse anealing protocol we initialize the system at $s=1$ with a classical state loaded into the qubits.  We then reduce $s$ quickly, bringing the system to the quantum model we want to simulate.  We let this model relax for a fixed evolution time of $\SI{65}{ms}$ before quickly quenching $s$ to $1$.

Each call to the processor consists of $50$ reverse anneals. A single initial state (clock or random) is loaded for the first anneal, and each subsequent anneal is initialized with the final state of the previous anneal.  This allows us to perform {\em quantum evolution Monte Carlo} to estimate equilibrium statistics of the target system.  From each processor call, we discard the first $25$ of $50$ samples as described in the main text.  Data shown in the main text represent 480 processor calls (with the exception of Fig.~\ref{fig:orderparameter}c, which represents 720 processor calls) divided equally between clock initial state and random intial state, each of which is further divided equally between two physical embeddings of the lattice (Supplemental \ref{sec:embedding}).

We make two adjustments to the QA schedule to correct deviation between the rf-SQUID flux qubits and the two-level Ising spins \cite{Harris2010}.  First we shift the schedule according to the deviation between the spectra of 2-level and 8-level models in a 4-SQUID FM-coupled chain; the result is a nonlinear shift in $s$ of approximately $0.02$.  Second we adjust the programmed coupling strengths to compensate for the background susceptibility $\chi_b$\cite{Albash2015}.  This effect results in a next-nearest-neighbor coupling between any two qubits that are coupled to a third qubit.  The magnitude of $\chi_b$ is a function of $s$ and is determined by independent measurements.  We compensate for this effect by adjusting couplings to homogenize effective AFM couplings and effective FM couplngs (see Supplemental \ref{sec:qa}).

We maintain adjustments to the general-purpose calibration to minimize the effect input-specific crosstalks.  This involves maintaining fine flux-bias offsets and coupling energy adjustments that respect the spin-flip and rotational symmetries in the cylindrical lattice (see Supplemental \ref{sec:shim}).  These symmetries are independent of temperature and transverse field, so these adjustments do not result in overfitting.

\subsection*{Classical Monte Carlo methods}\label{meth:qmc}

For all classical QMC experiments in this work we use continuous-time path-integral Monte Carlo \cite{Rieger1999} with Swendsen-Wang\cite{Swendsen1986} updates.  For a given spin system, a model is parameterized by $T$ and $\Gamma$, where $J=1$.  Here we describe methods for simulations and results described in the main text; further experiments and methods are discussed in the Supplemental Material.  We employ parallel tempering using a series of Hamiltonians that vary in $T$, with either $\Gamma$ or $\Gamma/T$ fixed.  We also employ a specialized chain update, in which every second Monte Carlo sweep attempts flipping of all spins in a four-qubit FM-coupled chain, rather than one spin at a time.  We find that this speeds up convergence by up to three orders of magnitude.

Convergence for a set of Hamiltonians is determined via standard error on $\langle m^2\rangle$ and convergence of absolute sublattice magnetizations to at most $0.04$ from a fully-magnetized classical clock initial state, which is then checked for consistency with the same experiment given a random initial state.  Convergence of the Binder cumulant and moments of the order parameter are tested with respect to initialization in several different types of classical ground state.  Error is determined self-consistently between five independent runs.  Pseudospin correlations are computed from $5120$ classical states projected from QMC in ten independent runs.  Experimental lengths are up to $2^{22}$ Monte Carlo sweeps.

\subsection*{Statistical methods}\label{meth:statistical}

All error bars show 95\% bootstrap confidence intervals over 1000 bootstrap samples.  For QA results in Fig.~\ref{fig:orderparameter}b and Fig.~\ref{fig:correlation}, independent confidence intervals are generated for the Monte Carlo estimators with ordered initial state and random initial state, and the union of these intervals is shown.  For every data point analyzed in Fig.~\ref{fig:orderparameter}c we also show the equivalent data points under reversal of all spins and/or rotation (but not flipping) of the cylinder.  Thus we show 27,000 effective samples for QA and 30,720 effective samples for QMC.

\subsection*{Phase diagram of square-octagonal lattice}\label{meth:phasediagram}

We determined the phase diagram of the square-octagonal lattice based on the statistics of $L\times L$ instances with toridal boundary conditions with $L$ between $3$ and $21$.  Following Isakov and Moessner\cite{Isakov2003} we use two methods to determine the KT critical temperatures: the critical exponent of $\braket m$ and universal scaling collapse.  To investigate the quantum critical point at $\Gamma_c\approx 1.76$, we show a crossing and universal collapse for the Binder cumulant for $L$ between $3$ and $15$ with inverse temperature $\beta(L,\Gamma) = 3.5L/\Gamma$.  Our results are further discussed and presented in Supplemental \ref{sec:phasediagram}.

\widetext
\pagebreak
\clearpage
\begin{center}
\makeatletter
\textbf{\large Supplemental Materials: \@title}
\makeatother
\end{center}
\setcounter{equation}{0}
\setcounter{figure}{0}
\setcounter{table}{0}
\setcounter{section}{0}
\setcounter{page}{1}
\makeatletter
\renewcommand{\thesection}{S-\Roman{section}}
\renewcommand{\theequation}{S\arabic{equation}}
\renewcommand{\thefigure}{S\arabic{figure}}

\section{Pseudospin phase and topological features in the TFIM}\label{sec:orderparameter}

In the TFIM, the triangular AFM lattice and fully-frustrated square-octagonal lattice admit a mapping to the 2D XY model that is essential to the topological phenomena we observe.  Here we provide a description of this mapping, which has its origins in the study of the stacked magnet\cite{Blankschtein1984a} and has since been described in the quantum case \cite{Moessner2000,Moessner2001,Isakov2003,Jiang2005,Wang2017}.  We begin with a description of the XY model and the upper KT phase transition intersected by the QA schedule.

\subsection{Two-dimensional XY model and the vortex unbinding transition}

\begin{figure}
\begin{tikzpicture}
\node[anchor=north west,fill=white] (1) at (0,0)
{\includegraphics[scale=.6]{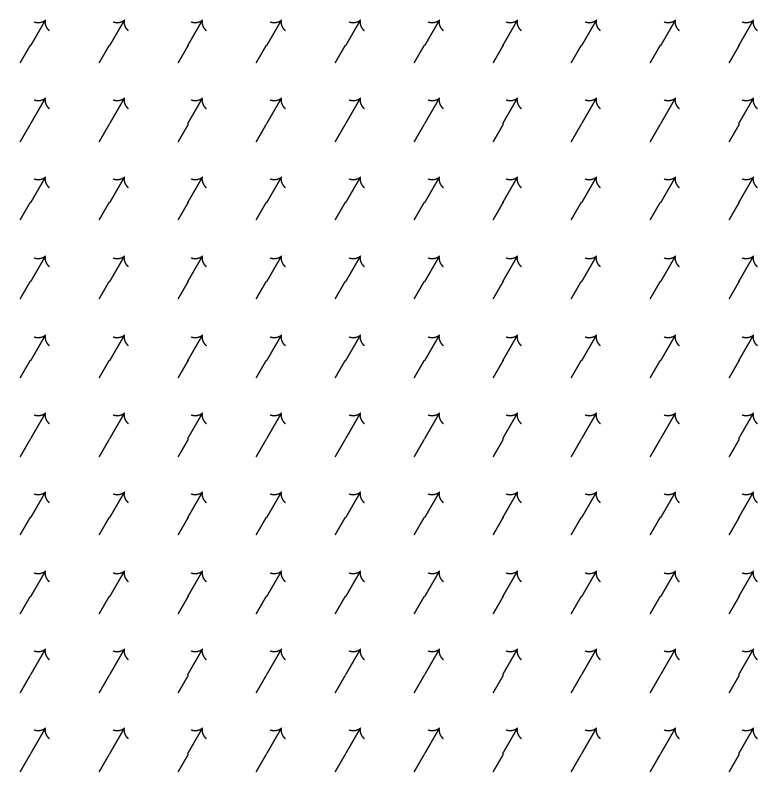}};
\node[anchor=north west,fill=white] (2) at (6,0)
{\includegraphics[scale=.6]{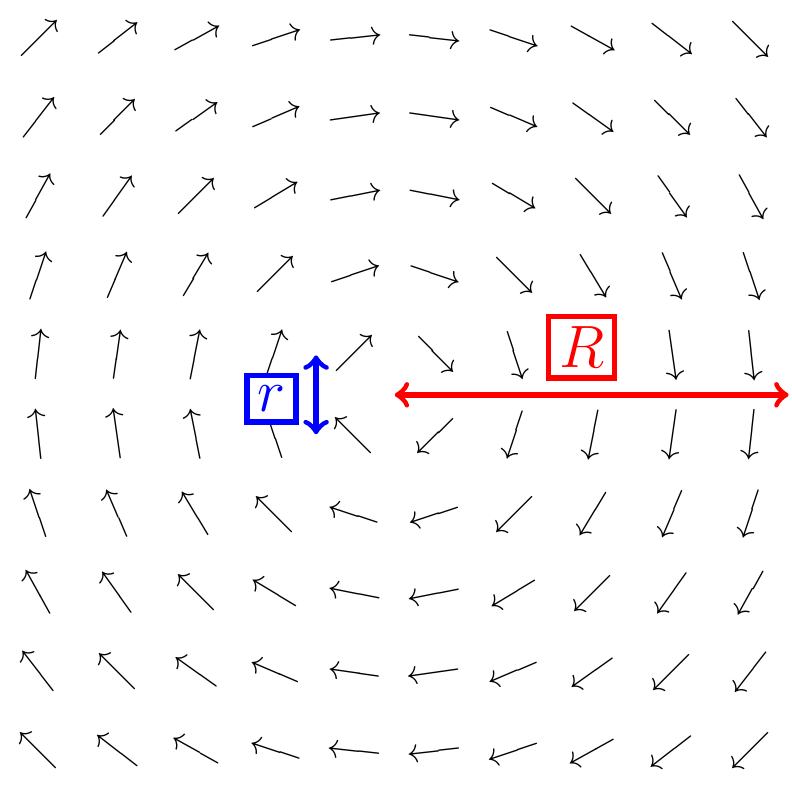}};
\node[anchor=north west,fill=white] (3) at (12,0)
{\includegraphics[scale=.6]{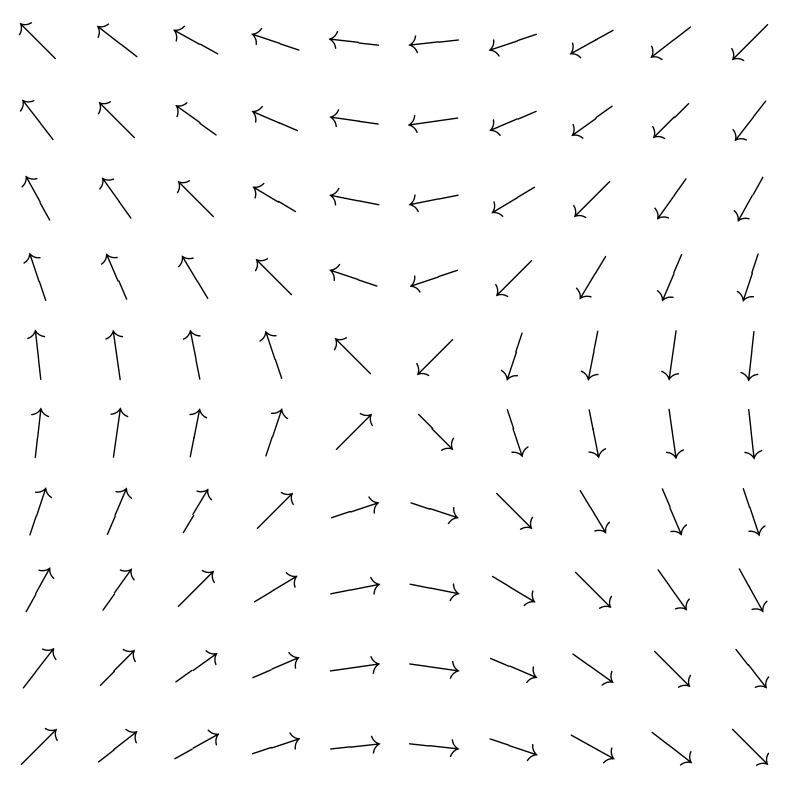}};
\node (a) at (0,0) {{\bfseries\sffamily\large a}};
\node (b) at (6,0) {{\bfseries\sffamily\large b}};
\node (c) at (12,0) {{\bfseries\sffamily\large c}};
\end{tikzpicture}
\caption{{\bf Topological features in the 2D XY model.}  {\bf a}, The ground state, in which all rotors have equal angle $\theta$, has continuous $U(1)$ symmetry, as a universal rotation of $\theta$ will not change the energy of the system. {\bf b--c}, A vortex or antivortex ({\bf b} and {\bf c} respectively) is a point around which the angle winds (rotates completely) clockwise or anticlockwise respectively when traversing a closed loop around the point in a clockwise direction.  Both the energy and the entropy of an isolated vortex or antivortex region of radius $R$ is proportional to $\log{(R/r)}$, where $r$ is the lattice spacing.}\label{fig:xy}
\end{figure}

The 2D XY model describes an interaction of classical spins.  Each spin is a unit vector described by an angle $\theta_i$.  The interaction between spins is given by the XY Hamiltonian
\begin{equation}\label{eq:xy}
H = -J_{\text{XY}}\sum_{\braket{i,j}}\cos{(\theta_i-\theta_j)},
\end{equation}
where the sum is taken over all coupled pairs of spins.  The ground state of this system is one in which all spins are aligned (see Fig.~\ref{fig:xy}); note that the system described by (\ref{eq:xy}) has continuous $O(2)$ or $U(1)$ symmetry.

In a ground state, the angle $\theta$ does not change as we traverse a closed path in the plane in a clockwise direction.  Given a perturbation of a ground state, $\theta$ may fluctuate along this path but will not wind in a full rotation.  A clockwise or anticlockwise winding is known as a vortex or an antivortex respectively; these topological defects imply an excitation that diverges with the radius $R$ of the encircling path:
\begin{equation}\label{eq:evortex}
E_{\text{vortex}} = \pi J_{\text{XY}}\log\tfrac R r,
\end{equation}
where $r$ is the lattice spacing.  The entropy $S_{\text{vortex}}$ also grows proportionally to $\log\tfrac Rr$ since there are $(R/r)^2$ possible locations of the vortex in an area of $R^2$.  A competition of these two terms leads to a sign-variable expression for the free energy contribution of a vortex:
\begin{equation}\label{eq:fvortex}
\Delta F = E_{\text{vortex}} - TS_{\text{vortex}}  = (\pi J_{\text{XY}} - 2k_BT)\log\tfrac R r,
\end{equation}
The critical vortex unbinding temperature $T_2$ at which the Kosterlitz-Thouless phase transition occurs is given by the sign change of $\Delta F$:
\begin{equation}\label{eq:tc}
T_2 = \tfrac \pi {2k_B}  J_{\text{XY}}.
\end{equation}
Above this temperature, the free energy favors isolated vortices; below this temperature vortices and antivortices are attracted to one another and appear in bound pairs.

\subsection{Pseudospins in geometrically frustrated quantum magnets}

\begin{figure}
\begin{tikzpicture}
\node[anchor=south west,fill=white] (1) at (0,0)
{\includegraphics[scale=.9]{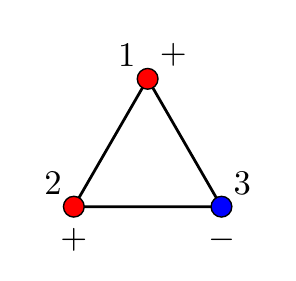}};
\node[anchor=south west,fill=white] (2) at (3,0)
{\includegraphics[scale=.9]{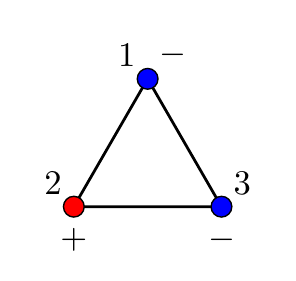}};
\node[anchor=south west,fill=white] (3) at (6,0)
{\includegraphics[scale=.9]{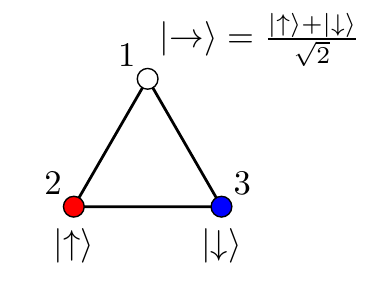}};
\node (a) at (1,3) {{\bfseries\sffamily\large a}};
\node (b) at (4,3) {{\bfseries\sffamily\large b}};
\node (c) at (7,3) {{\bfseries\sffamily\large c}};
\end{tikzpicture}\vspace{-.25cm}
\caption{{\bf Floppy spin aligns with the transverse field.}  {\bf a--b}, Two of six degenerate classical ground states differ by a flip of spin $1$.  {\bf c}, With the addition of a perturbative transverse field $\Gamma$, spin $1$ aligns with the transverse field in a symmetric superposition $\ket\rightarrow = (\ket \uparrow + \ket \downarrow)\sqrt{2}$, reducing the energy from $-J$ to $-J-\Gamma$.}\label{fig:triangle1}
\end{figure}

The mapping from the TFIM to the XY model arises from the interaction of frustration and quantum fluctuations.  We first consider the effect of a perturbative transverse field on a single AFM triangle.  In the classical case, the AFM triangle has six ground states.  In each, one of two {\em floppy} spins can be flipped without changing the energy.  Fig.~\ref{fig:triangle1} shows two such states differing by a flip of spin $1$.  The addition of a perturbative transverse field $\Gamma$ lifts this degeneracy, allowing a superposition of these two states in which spin $1$ aligns with the transverse field.  We denote this superposition by $\ket\rightarrow = (\ket \uparrow + \ket \downarrow)\sqrt{2}$ and note that this lowers the energy from $-J$ to $-J-\Gamma$.

\begin{figure}
\begin{tikzpicture}
\node[fill=white] (3) at (0,0)
{\includegraphics[scale=.9]{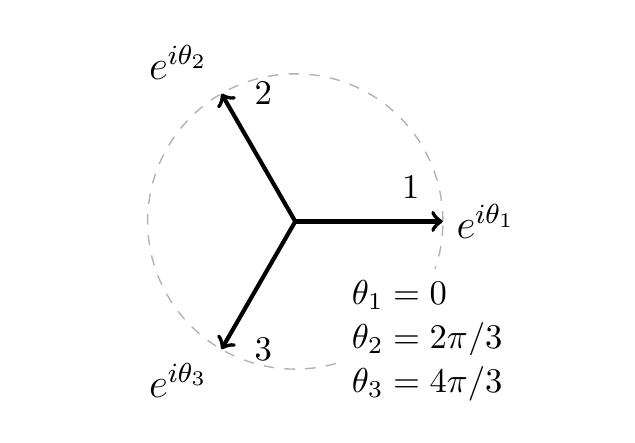}};
\node[fill=white] (1) at (5,0)
{\includegraphics[scale=.9]{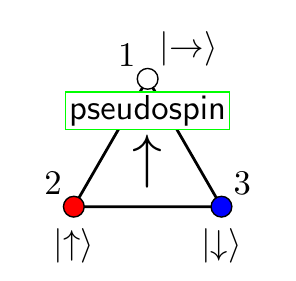}};
\node[fill=white] (3) at (10,0)
{\includegraphics[scale=.9]{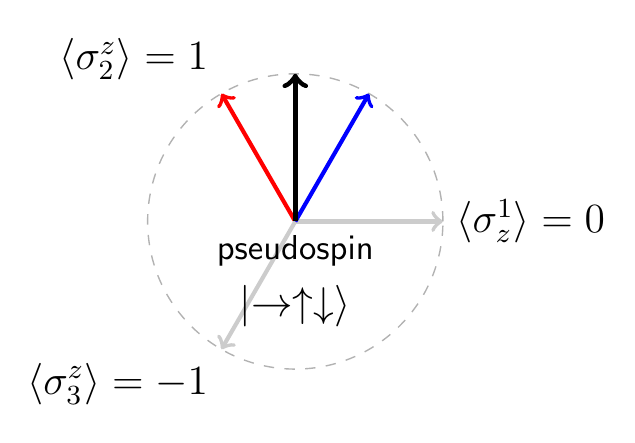}};
\node[fill=white] (3) at (5,-4.5)
{\includegraphics[width=7cm]{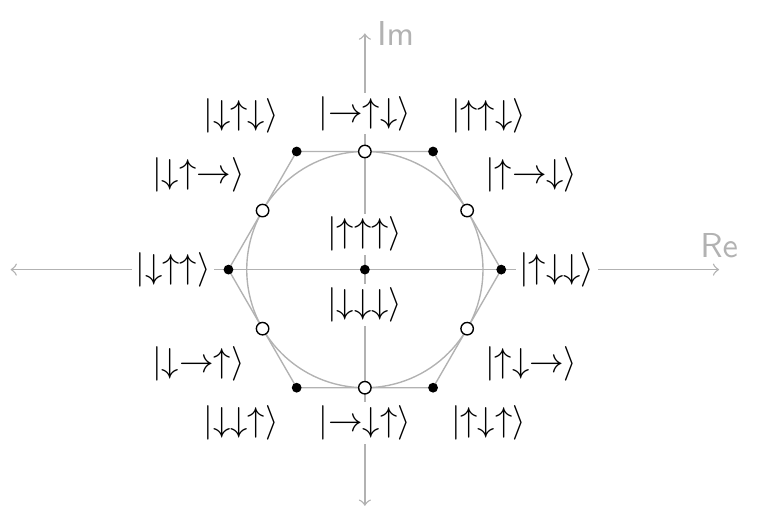}};
\node (a) at (0,2) {{\bfseries\sffamily\large a}};
\node (b) at (5,2) {{\bfseries\sffamily\large b}};
\node (c) at (10,2) {{\bfseries\sffamily\large c}};
\node (b) at (2,-3) {{\bfseries\sffamily\large d}};
\end{tikzpicture}
\caption{{\bf Pseudospins from classical and quantum states.} {\bf a}, Each pseudospin $\psi_j$ is determined as a linear combination of basis vectors with weights given by $\sigma^z$ operators. {\bf b--c}, the pseudospin of the clock state $\ket{\rightarrow\uparrow\downarrow}$ is $e^{i\pi/2}$. {\bf d} The six-fold degenerate perturbative quantum ground states (white, with magnitude $1$) and six-fold degenerate classical ground states (black, with magnitude $2/\sqrt{3}$) admit twelve clock pseudospins in the complex plane.  Also shown are the classical excited states $\ket{\uparrow\uparrow\uparrow}$ and $\ket{\downarrow\downarrow\downarrow}$, which have pseudospin $0$ and therefore correspond to a vortex or antivortex.} \label{fig:triangle2}
\end{figure}

To first order perturbation, this quantum ground state is six-fold degenerate, with the three spins taking values $\ket\uparrow$, $\ket\downarrow$, $\ket\rightarrow$ in any permutation.  We map these six states to the complex plaquette pseudospins $\psi_j$ as described by (\ref{eq:pseudospin}): $\psi_j = \braket{\sigma_1^z} + \braket{\sigma_2^z}e^{2\pi i /3} + \braket{\sigma_3^z}e^{4\pi i /3}$.  This is shown in Fig.~\ref{fig:triangle2} for the state $\ket{\rightarrow\uparrow\downarrow}$.  Pseudospins for six quantum and six classical states are shown in Fig.~\ref{fig:triangle2}b.  Also shown are the classical excited states $\ket{\uparrow\uparrow\uparrow}$ and $\ket{\downarrow\downarrow\downarrow}$, which have $\psi_j=0$.  As vanishing points of the pseudospin phase, these magnetized triangles correspond to vortices and antivortices in the lattice.

\begin{figure}
\begin{tikzpicture}
\node[anchor = north west,fill=white] (1) at (0,0)
{\includegraphics[scale=1]{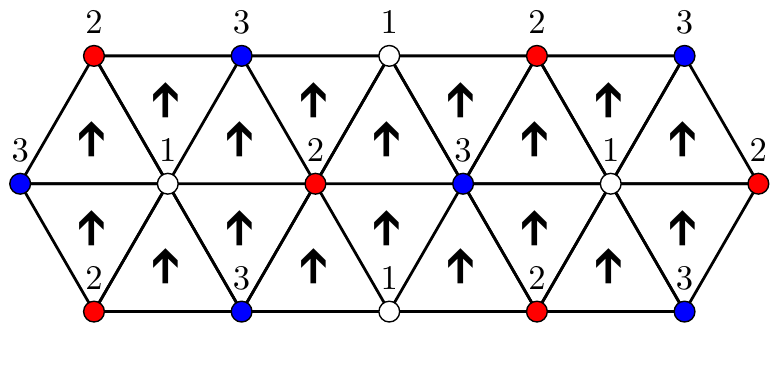}};
\node[anchor = north west,fill=white] (3) at (9,0)
{\includegraphics[scale=1]{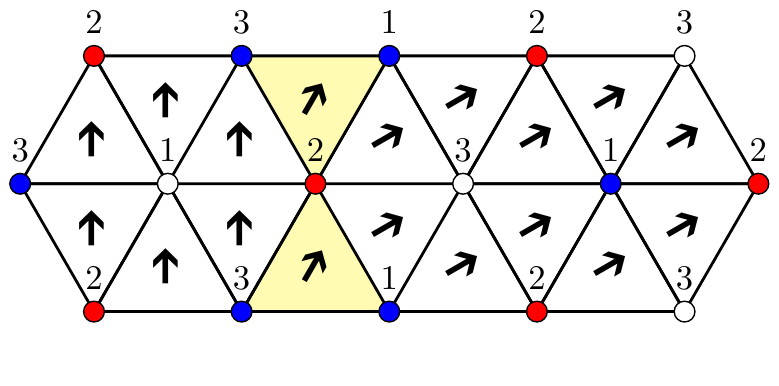}};
\node (a) at (0,-.5) {{\bfseries\sffamily\large a}};
\node (b) at (9,-.5) {{\bfseries\sffamily\large b}};
\end{tikzpicture}\vspace{-.25cm}
\caption{{\bf Fluctuations in pseudospin phase.}  {\bf a}, A clock state can be tiled across the lattice with uniform pseudospin phase.  To first order perturbation this results in the system having six-fold ground state degeneracy, as in the case of a single AFM triangle.  {\bf b}, A rotation of the pseudospins leads to triangles with no transverse-field-aligned spin, giving an excitation proportional to $\Gamma$ that is analogous to the XY model.}\label{fig:triangle3}
\end{figure}

As shown in Fig.~\ref{fig:triangle3}a, the trianglular AFM lattice can be tiled with a single clock state, resulting in a pseudospin that is constant across the lattice.  This tiling naturally divides the sites of the lattice into three sublattices, which we use to determine the complex order paramater $\psi$ as in (\ref{eq:orderparameter}).  The introduction of a twist in the pseudospin phase, as shown in Fig.~\ref{fig:triangle3}b, leads to a state where two of the triangles contain no spin aligned with the transverse field.  This excitation can be described in terms of an effective XY model on the dual lattice, with effective XY coupling proportional to $\Gamma$.  This is consistent with the linear scaling of $T_2$ in $\Gamma$ seen in the small-$\Gamma$ limit\cite{Korshunov2012}.

\begin{figure}
\begin{tikzpicture}
\node[anchor=south east,fill=white] (vectorfield) at (-.2,6.5)
{\includegraphics[width=8cm]{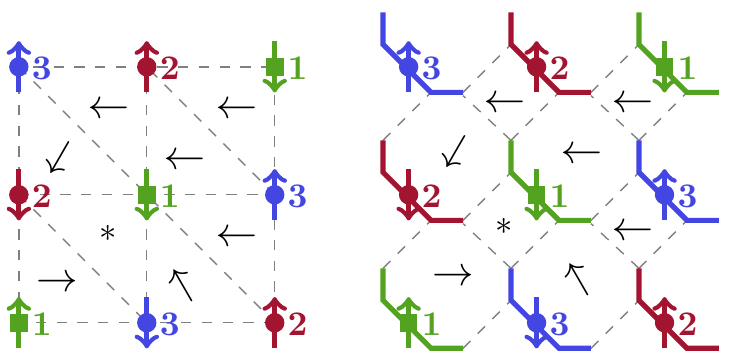}};
\node (a) at (-8.5,10.4) {{\bfseries\sffamily\large a}};
\node (b) at (-4.4,10.4) {{\bfseries\sffamily\large b}};
\end{tikzpicture}
\caption{{\bf Complex order parameter in the transverse field Ising model.}  {\bf a--b}, Each spin in the triangular ({\bf a}) and square-octagonal ({\bf b}) belongs to one of three sublattices, indicated by marker shape and color.  The local order parameter term due to that spin is the Ising spin ($\pm 1$) multiplied by the sublattice's basis vector; i.e., $1$, $e^{i2\pi/3}$, or $e^{i4 \pi/3}$ depending on the sublattice.  Averaging these terms for the spins of triangle $j$ in the triangular lattice and normalizing gives the value $\psi_j$ for each plaquette.  These values are marked by arrows in the plaquettes, with a vanishing point marked with $*$.}\label{fig:triangle_op}
\end{figure}

Fig.~\ref{fig:triangle_op} shows a vortex in the triangular lattice, and in the corresponding state in the square-octagonal lattice.  To compute the pseudospin of a plaquette in the square-octagonal phase we use all 12 spins in the three FM-coupled chains intersecting the plaquette, whether it is a square or an octagon.

\clearpage

\section{Quantum processor and experimental methods}\label{sec:qa}

The quantum annealing processor used in this work is a D-Wave 2000Q system operated in a non-annealing mode.  Operating temperatures ranging between $8.4\pm 0.2$ mK and $21.4\pm 0.2$ mK are measured with a cryostat thermometer and independently verified using single-qubit susceptibility measurements.

\subsection{Annealing schedule and reverse annealing protocol}\label{sec:protocols}
\begin{figure}
\begin{tikzpicture}\sf
\node[anchor=north] (mversuss) at (0,0)
{\includegraphics[scale=.82]{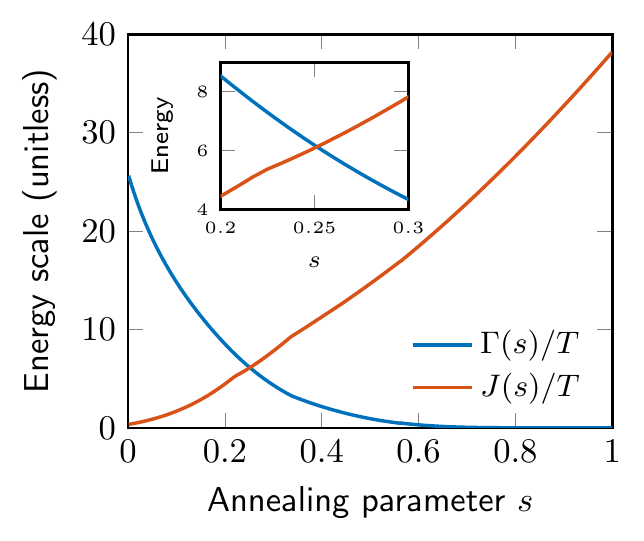}};
\node[anchor=north] (mversuss) at (6,0)
{\includegraphics[scale=.82]{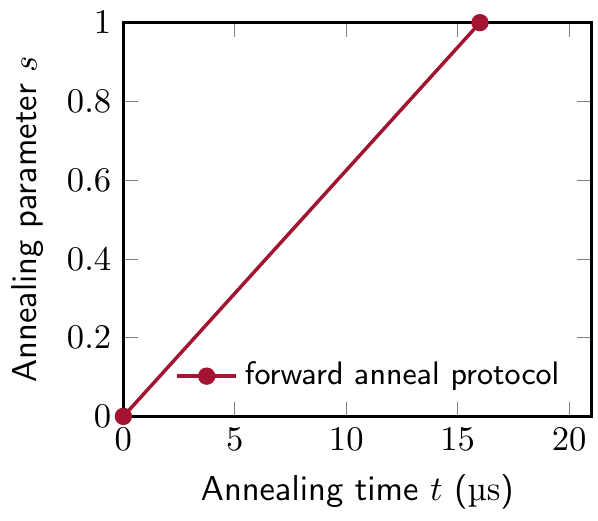}};
\node[anchor=north] (mversuss) at (12,0)
{\includegraphics[scale=.82]{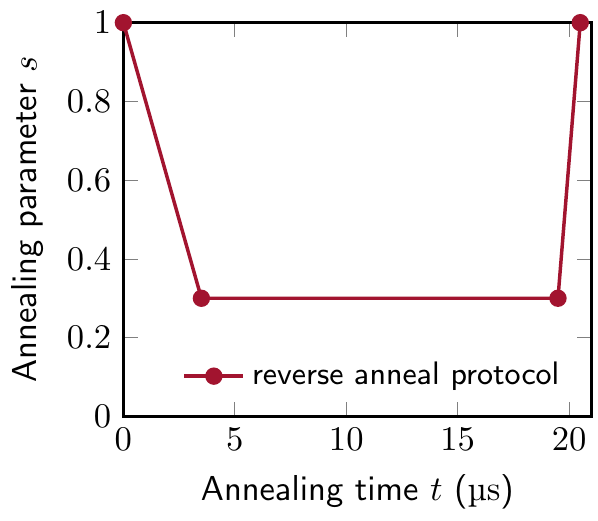}};

\draw[<-,thick]  (10.8,-.5) -- (11.2,-.75)  node[right] {input};
\draw[<-,thick]  (14.2,-.5) -- (13.8,-.7)  node[left] {readout};
\draw[<-,thick]  (7.2,-.5) -- (6.8,-.7)  node[left] {readout};
\node (a) at (-2.6,-0.2) {{\bfseries\sffamily\large a}};
\node (b) at (3.3,-0.2) {{\bfseries\sffamily\large b}};
\node (c) at (9.4,-0.2) {{\bfseries\sffamily\large c}};
\end{tikzpicture}\vspace{-.25cm}
\caption{{\bf Annealing schedule and protocols.}  {\bf a}, Unitless energy scales for transverse field $\Gamma$ and Ising couplings $J$ as a function of annealing parameter $s$, compared with temperature $T=\SI{8.4}{mK}$.  {\bf b}, Standard $\SI{16}{\micro s}$ forward anneal protocol given as $s(t)$, which increases linearly from $0$ to $1$.  {\bf c}, $\SI{16}{\micro s}$ evolution in a reverse annealing protocol, where $s$ drops from $1$ to $s=0.3$ over $\SI{3.5}{\micro s}$, dwells at $s$ for $\SI{16}{\micro s}$, and quenches to $1$ over $\SI{1}{\micro s}$.  Experiments are performed with $2^{16}\SI{}{\micro\second}$ ($\SI{65}{ms}$) dwells at target model $s$.  }\label{fig:protocols}
\end{figure}

We focus on the region of the annealing schedule between $s=0.20$ and $s=0.30$ (Fig.~\ref{fig:protocols}a).  In quantum annealing, the annealing parameter $s$ is typically increased smoothly from $0$ to $1$, bringing the system from a single-well superposition at $s=0$ to a low-temperature classical system at $s=1$ (Fig.~\ref{fig:protocols}b).  In our experiments we take a different approach, using a {\em reverse annealing} protocol.  In this protocol we initialize the system at $s=1$ with a classical state loaded into the qubits.  We then reduce $s$ quickly, bringing the system to the quantum model we want to simulate.  We let this model relax for a fixed evolution time of $\SI{65}{ms}$ before quickly quenching $s$ to $1$ (Fig.~\ref{fig:protocols}c).

Each call to the processor consists of $50$ reverse anneals.  A single initial state (clock or random) is loaded for the first anneal, and each subsequent anneal is initialized with the final state of the previous anneal (Fig.~\ref{fig:protocols}c).  This allows us to perform {\em quantum evolution Monte Carlo} to estimate equilibrium statistics of the target system.  From each processor call, we discard the first $25$ of $50$ samples as described in the main text.

\begin{figure*}
\begin{tikzpicture}\setlength{\figurewidth}{6cm}\setlength{\figureheight}{4.5cm}\sf
\node[anchor=north] (mversuss) at (0,0)
{\includegraphics[scale=.82]{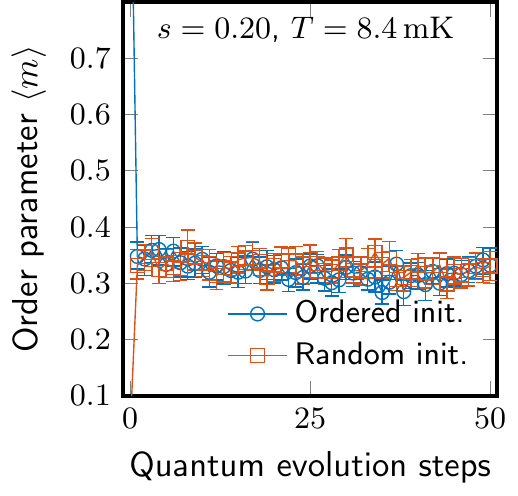}};
\node[anchor=north] (mversuss) at (5,0)
{\includegraphics[scale=.82]{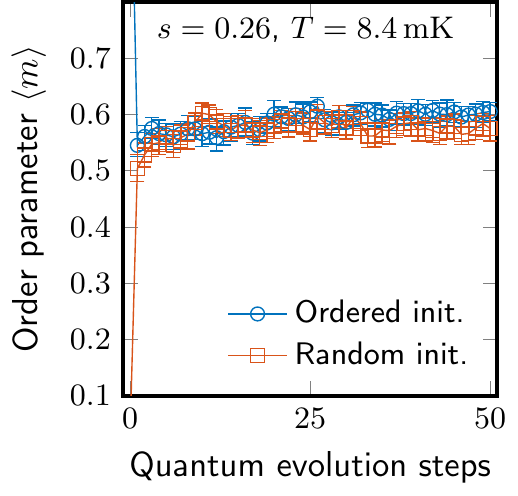}};
\node[anchor=north] (mversuss) at (10,0)
{\includegraphics[scale=.82]{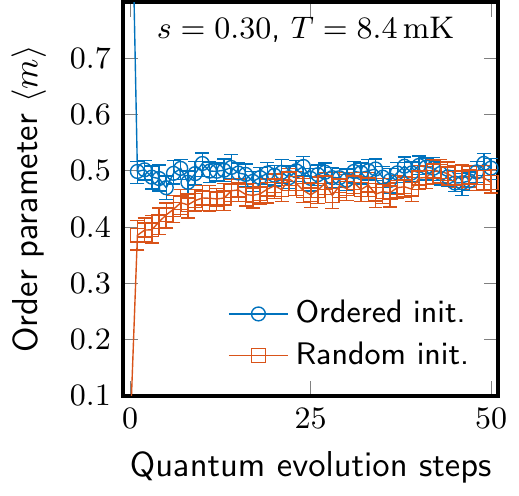}};
\node[anchor=north] (mversuss) at (0,-5)
{\includegraphics[scale=.82]{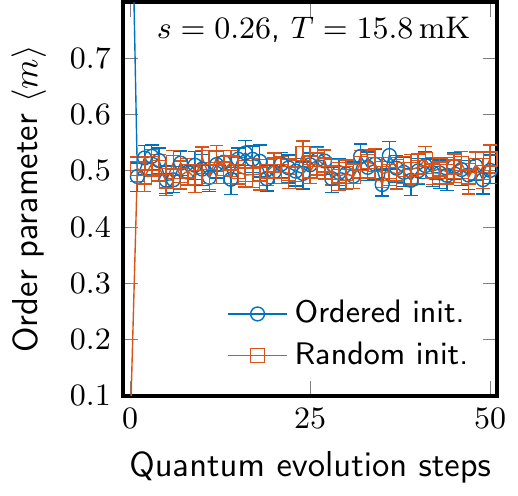}};
\node[anchor=north] (mversuss) at (5,-5)
{\includegraphics[scale=.82]{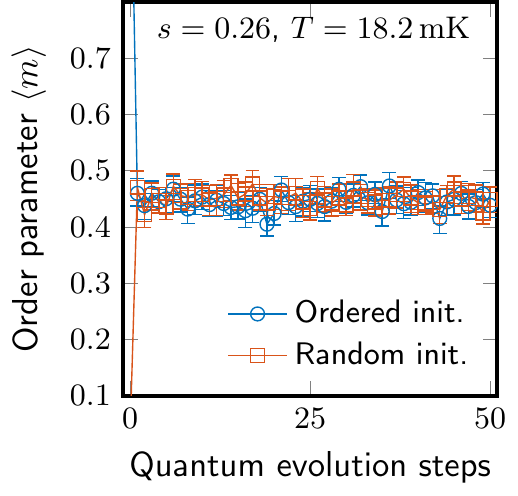}};
\node[anchor=north] (mversuss) at (10,-5)
{\includegraphics[scale=.82]{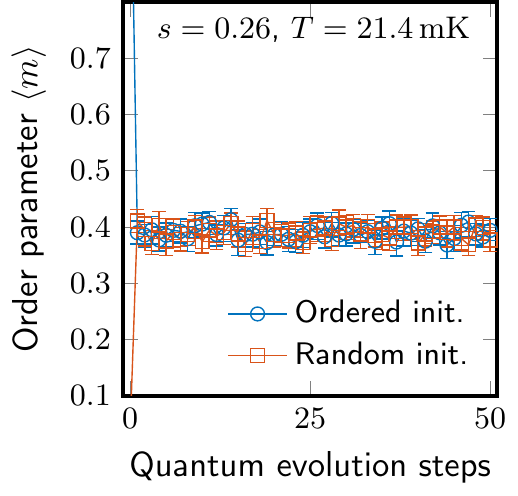}};
\end{tikzpicture}\vspace{-.25cm}
\caption{{\bf Convergence of quantum evolution Monte Carlo.}  Our quantum evolution Monte Carlo sampling approach involves making a chain of 50 reverse annealing evolutions starting from an ordered state and a random state.  At $\SI{8.4}{mK}$, cooling of around $\SI{0.1}{mK}$ is observed during the first 25 evolutions.  This cooling lowers $m$ at $s=0.26$, as the lower temperature slows evolution during the $\SI{1}{\micro\second}$ quench.  The same cooling increases the order parameter at $s=0.26$, where the lower temperature results in evolution of a more ordered model.  Shown are experimental results for increasing values of $s$ (top row) and for $s=0.26$ at increasing temperatures (bottom row).  At $s=0.30$, $T =\SI{8.4}{mK}$, evolution is slow and the estimates of $\langle m\rangle$ are far from the equilibrium value of $\langle m\rangle$ after a single $\SI{65}{ms}$ evolution.  Spin-bath polarization, which biases the estimators towards the initial condition at low temperatures,  disappears as temperature is increased.}\label{fig:markovchain}
\end{figure*}

When $\Gamma$ is small, the rf-SQUID flux qubits faithfully model two-level systems.  Since our experiments are performed in a region of the schedule where $\Gamma$ is large, behavior deviates significantly from a two-level system but is well described by an eight-level model\cite{Harris2010}.  To account for this we make two adjustments.  First we shift the schedule according to the deviation between the spectra of 2-level and 8-level models in a 4-SQUID FM-coupled chain.  Second we adjust the programmed coupling strengths to compensate for the background susceptibility effect described in Supplemental \ref{sec:chi}.

\subsection{Embedding the square-octagonal lattice}\label{sec:embedding}

\begin{figure}
\begin{tikzpicture}
\node[anchor=west] (state2) at (0,0)
{\includegraphics[width=6cm]{embedded_lattice.pdf}};
\node[anchor=west] (state1) at (7.5,0)
{\includegraphics[width=7cm]{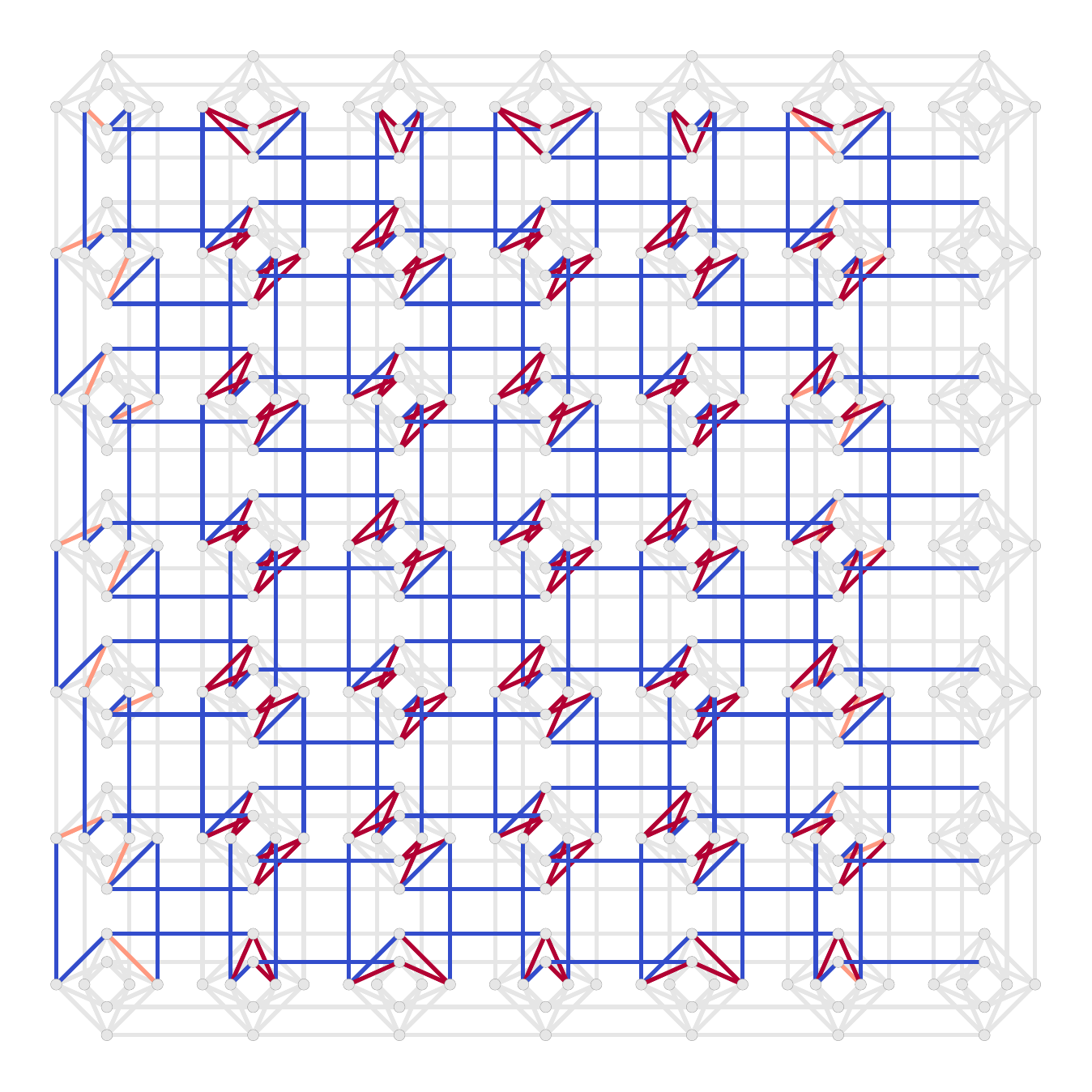}};
\node[anchor=west] (state1) at (0,-11)
{\includegraphics[width=14cm]{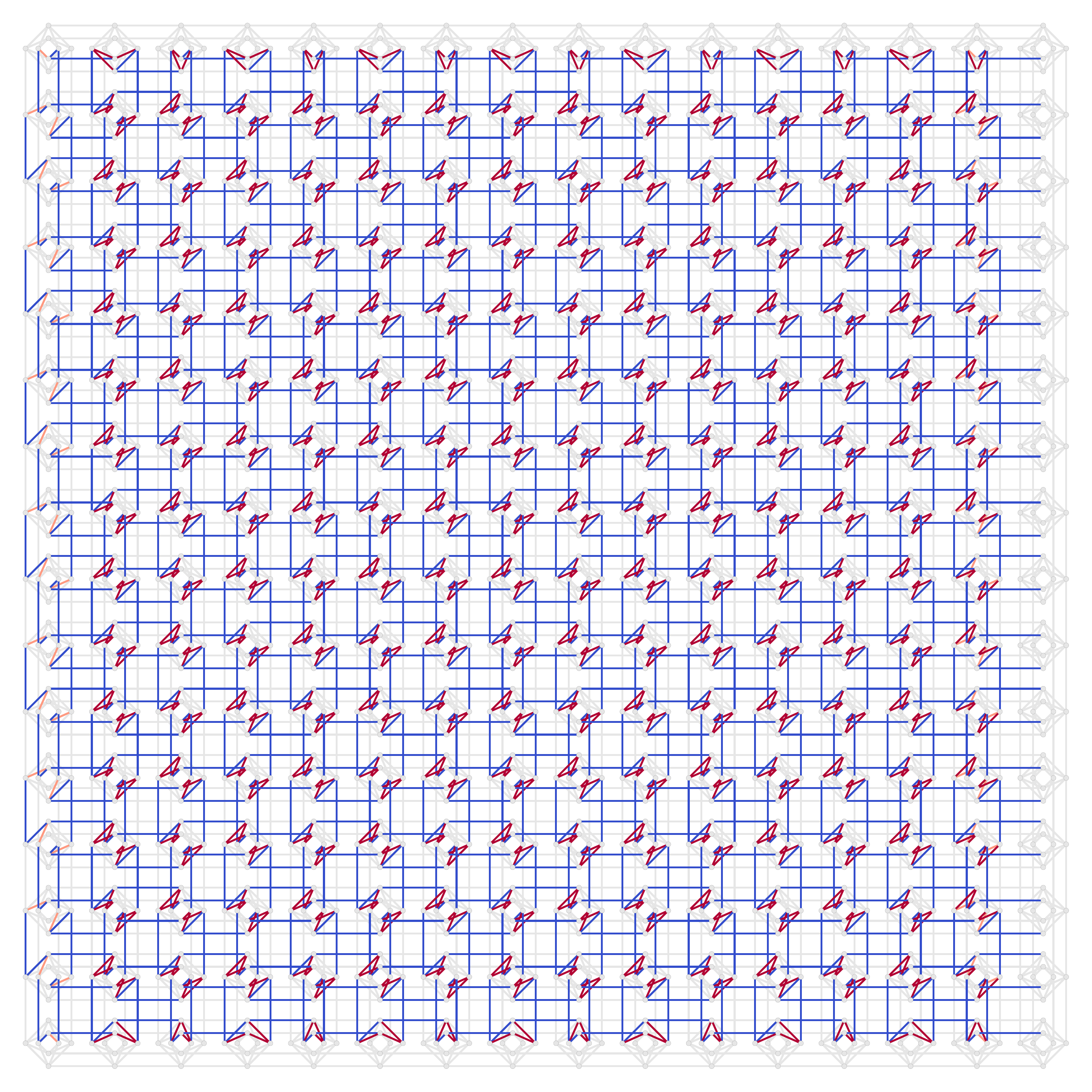}};
\node (a) at (0,2.5) {{\bfseries\sffamily\large a}};
\node (b) at (7.5,2.5) {{\bfseries\sffamily\large b}};
\node (a) at (0,-4.5) {{\bfseries\sffamily\large c}};
\end{tikzpicture}\vspace{-.25cm}

\caption{{\bf Embedding a cylindrical lattice into the qubit connectivity graph.} The cylindrical square-octagonal lattice with length $L=6$ ({\bf a}) is embedded into a region of the {\em Chimera} qubit connectivity graph ({\bf b}).  The embedding consists of two square sheets that are coupled together at the top and bottom.  The largest instance studied ({\bf c}) with $L=15$ uses 1,800 of the 2,048 qubits in the processor, with unused qubits along the outside boundary of the processor.  FM couplers are in blue and AFM couplers are in red.}\label{fig:embedding}
\end{figure}

The cylindrical topology of the square-octagonal lattice is realized as two square sheets coupled together at the top and bottom (Figure \ref{fig:embedding}).  In the processor used, the graph of available couplings is a $16\times 16$ grid of {\em Chimera} unit cells\cite{Bunyk2014}; only a $7\times 7$ block of cells is shown in Fig.~\ref{fig:embedding}b.  Each FM-coupled chain of four qubits consists of two qubits in one unit cell, and one qubit in each of two other unit cells.  If the size of the unit cell were doubled, it would be possible to embed a toroidal square-octagonal lattice with fully periodic boundary conditions.  The halving of the antiferromagnetic couplers on the open boundaries of the cylinder ensures that the classical ground state space reflects that of the toroidal lattice.

\subsection{Background susceptibility and compensation}\label{sec:chi}

The rf-SQUID flux qubits in the QA processor provide an imperfect implementation of two-level Ising spins.  The primary deviation from ideal is that early in the anneal, every Ising spin mediates a coupling between its neighboring spins.  This introduces next-nearest-neighbor couplings, meaning that for any spins $i,j,k$, there is an effective coupling term $\chi_bJ_{ij}J_{jk}$ added to $J_{ik}$;  the strength is denoted $\chi_b$, representing normalized {\em background susceptibility}: $\chi_b = M_{\rm AFM} \chi_q$ where $M_{\rm  AFM}$ is the maximum available AFM mutual inductance and $\chi_q$ is the physical qubit susceptibility.  The value of $\chi_b$ varies from $-0.07 \pm 0.02$ early in the anneal to $-0.03 \pm 0.01$ late in the anneal where single-spin dynamics freeze out, as measured by independent experiments. 

In the QA experiments, we compensate for $\chi_b$ based on the fact that FM couplers are seldom frustrated in the systems studied.  We consider the application of next-nearest-neighbor terms as a function $f$ applied to the coupling Hamiltonian $J_{\textrm{QA}}$ supplied to the processor, and seek a choice of $J_{\textrm{QA}}$ such that $f(J_{\textrm{QA}})$ approximates $\alpha J$ for some constant $\alpha>0$.  This gives us the following constraints on the classical Hamiltonians:
\begin{enumerate}
\item Any two chains have the same total coupling between them in $f(J_{\textrm{QA}})$ and $\alpha J$.  In the absence of boundary conditions, this coupling is $\alpha$.
\item If a chain has a single break (domain wall), the total strength of frustrated couplers is the same in $f(J_{\textrm{QA}})$ and $\alpha J$.  In the absence of boundary conditions, this coupling is $-1.8\alpha$.
\end{enumerate}
We use an iterative method to find $J_{\textrm{QA}}$ given an average AFM coupling strength $\rho$; it follows that $\alpha$ is a function of $\rho$ and $\chi_b$ (in turn, $\chi_b$ is a function of $s$). For $\rho = 0.95$ the value of $\alpha$ ranges monotonically from $1.32$ at $s=0.25$ to $1.13$ at $s=0.4$.

\subsection{Calibration refinement}\label{sec:shim}

The QA processor was calibrated as a general-purpose quantum annealer, with the goal of good performance across a variety of inputs.  Given our focus on one particular input type, we can improve performance by exploiting structure and symmetries in the input.  After compensating for background susceptibility as described above, we exploit symmetries in the lattice.

\begin{figure*}
\begin{tikzpicture}\setlength{\figurewidth}{6cm}\setlength{\figureheight}{4.5cm}\sf
\node[anchor=north] (mversuss) at (0,0)
{\includegraphics[scale=.82]{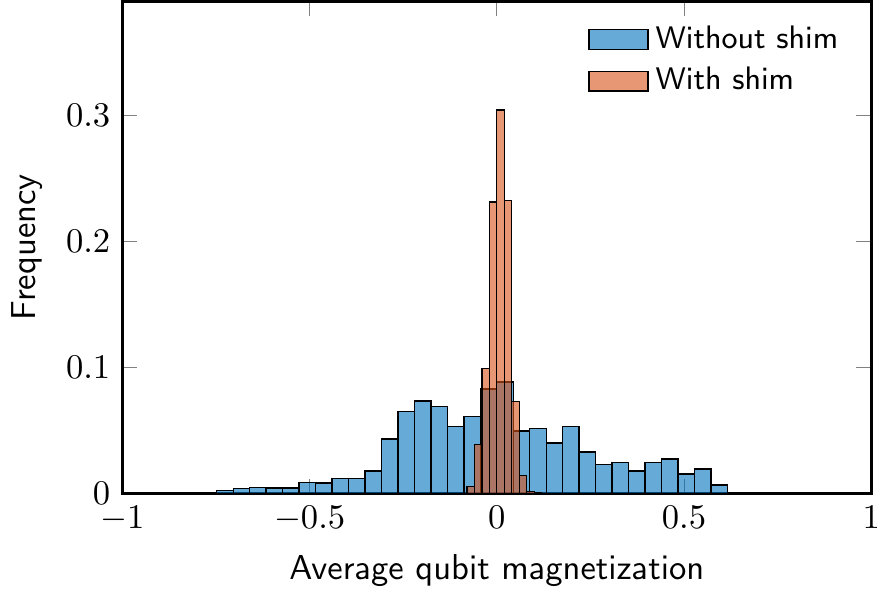}};
\node[anchor=north] (mversuss) at (8,0)
{\includegraphics[scale=.82]{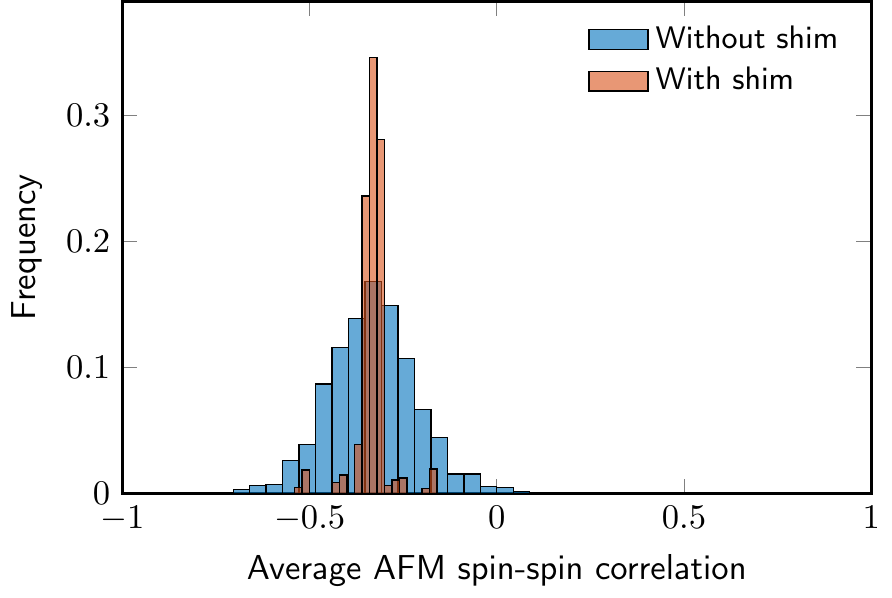}};
\node (a) at (-3.2,-.1) {{\bfseries\sffamily\large a}};
\node (b) at (4.8,-.1) {{\bfseries\sffamily\large b}};
\end{tikzpicture}\vspace{-.25cm}
\caption{{\bf Effect of calibration refinement.}  {\bf a--b}, Symmetries in the lattice dictate that each qubit should have an average magnetization of zero ({\bf a}) and that each AFM coupler should be frustrated with the same probability as other couplers in its rotational symmetry equivalence class of $2L$ couplers ({\bf b}).  We use a gradient-descent shim method to maintain degeneracy among qubits and equivalent couplers with flux offsets and small adjustments to specified coupling energies.  Data are taken for $L=15$, with 1,800 qubits and 1,290 AFM couplers, over 120 experiments with and without shim.  Both magnetization and correlation distributions are tightened significantly.  Outliers in shimmed spin-spin correlation are symmetry classes near the boundary of the lattice.}\label{fig:shim}
\end{figure*}

The first symmetry is spin-flip invariance: since there is no longitudinal field $h$, the expected magnetization of each spin in the absence of systematic biases should be zero.  To maintain this degeneracy we apply flux-offset biases to qubits in a gradient descent method, significantly improving the distribution of qubit magnetizations (Fig.~\ref{fig:shim}a).  The required offset is a function of annealing schedule, so we maintain an independent compensation for each schedule based on forward annealing results.

The second symmetry is rotational symmetry about the periodic dimension.  For lattice width $L$ (with $4\cdot L \cdot 2L$ spins), each coupler is in an equivalence class of $2L$ rotationally equivalent couplers that should be frustrated equiprobably.  More generally, two couplers should have the same statistics if there is a graph automorphism over the lattice that maps one to the other.  We make fine adjustments to individual coupling terms to tighten the frustration distributions of these equivalence sets (Fig.~\ref{fig:shim}b).

There are no further trivial symmetries in the square-octagonal lattice; both the triangular lattice with semi-open boundary and the square-octagonal lattice with fully periodic boundary have richer automorphism groups that might be exploited in future research.  The symmetries that we use for calibration refinement apply at every point in the phase diagram, so the calibration refinement used here does not result in overfitting.

\subsection{Spin-bath polarization}\label{sec:polarization}

The persistent current flowing in the qubit bodies during the QA protocol produces a magnetic field that can partially align or polarize the spin system.  For long evolution times, the polarized environment can produce sample-to-sample correlations, biasing the QA towards previously achieved spin configurations.  To reduce sample-to-sample correlations, we introduce a pause of $\SI{10}{ms}$ between anneals to give the spin-bath time to depolarize.  However in the reverse annealing protocol we start in a polarized state, and the QA system will be biased towards the initial state by polarization in the spin-bath.  This leads to the small systematic difference in estimators of $\langle m\rangle$ using a clock initial state versus a random initial state.

\section{Classical Monte Carlo methods}\label{sec:qmc}

For all classical QMC experiments in this work we use continuous-time path-integral Monte Carlo \cite{Rieger1999} with Swendsen-Wang\cite{Swendsen1986} updates.  For a given spin system, a model is parameterized by $T$ and $\Gamma$, where $J=1$.

When seeking equilibrium estimates of a model, we employ parallel tempering using a series of models that vary in $T$, with either $\Gamma$ or $\Gamma/T$ fixed.  We also employ a specialized chain update, in which every other Monte Carlo sweep attempts flipping of all spins in a four-qubit FM-coupled chain, rather than one spin at a time.  We find that this speeds up convergence by up to three orders of magnitude.

Convergence for a set of models is determined via standard error on $\langle m^2\rangle$ and convergence of absolute sublattice magnetizations to at most $0.04$ from a fully-magnetized clock initial state, which is then checked for consistency with the same experiment given a random initial state.  Convergence of the Binder cumulant and moments of the order parameter are tested  with respect to initialization in several different types of classical ground state.  Error is determined self-consistently between independent runs; error bars on QMC results are 95\% bootstrap confidence intervals.

\section{Phase diagram of square-octagonal lattice}\label{sec:phasediagram}

Here we provide evidence for the claimed phase diagram of the square-octagonal lattice.  We show critical behavior and give estimates of the upper and lower critical temperatures bounding the critical KT region in the phase diagram of the square-octagonal lattice, and give an estimate of the quantum critical point at $T=0$.  We follow the methods used by Isakov and Moessner\cite{Isakov2003} on the triangular lattice.  We study $L\times L$ lattices (on $4L^2$ spins) with fully periodic (toroidal) boundary conditions with $L$ ranging from $3$ to $21$.  Our results are presented in Figure \ref{fig:phasediagram}.

\subsection{KT critical temperatures}

For a range of transverse fields we use two methods to determine the lower ($T_1$) and upper ($T_2$) critical temperatures using properties of the 2D XY universality class.

\begin{figure}
\begin{tikzpicture}
\node[anchor=north east] (state1) at (0,.05)
{\includegraphics[scale = 0.82]{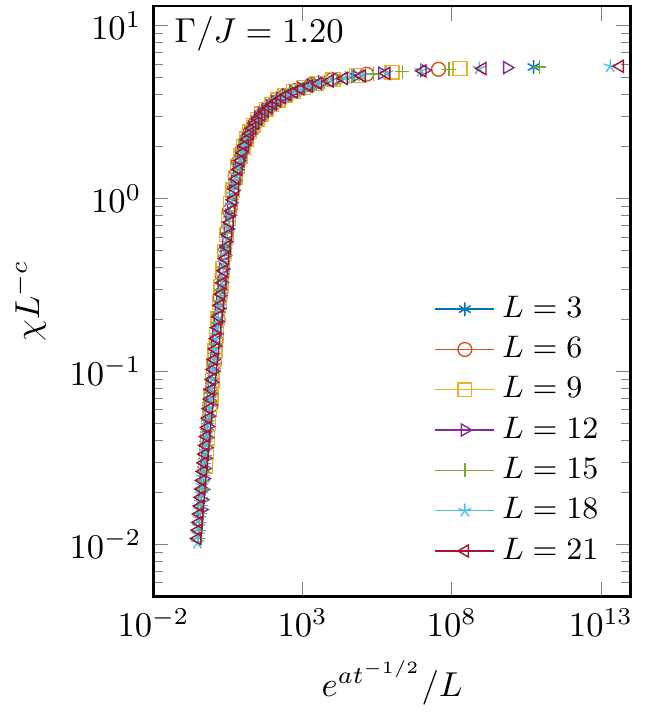}};
\node[anchor=north east] (state1) at (6.3,0)
{\includegraphics[scale = 0.82]{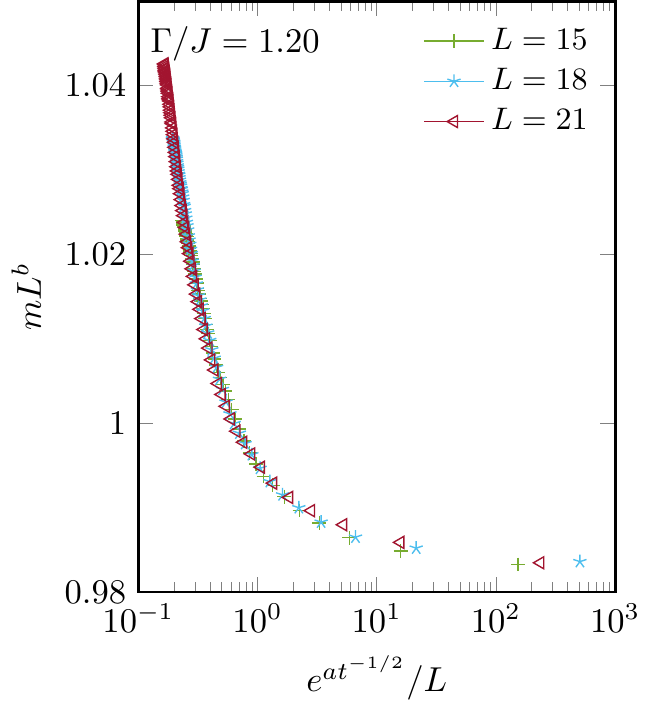}};
\node[anchor=north east] (phasediagram) at (12,0)
{\includegraphics[scale=.82]{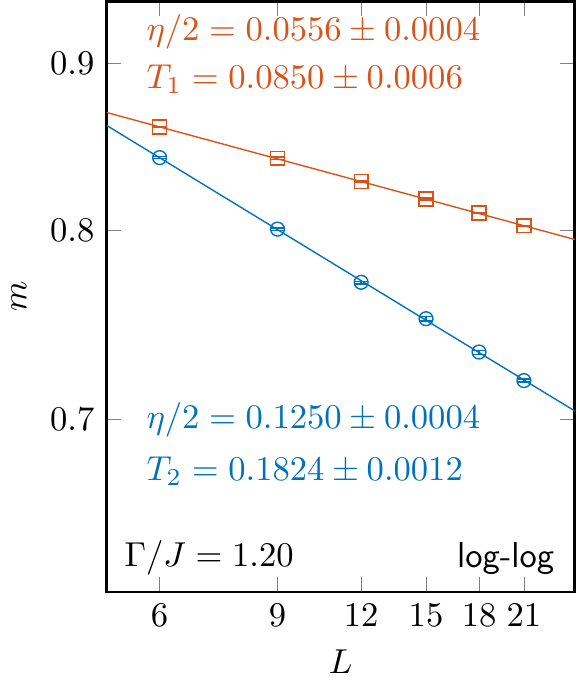}};
\node[anchor=north east]  at (3,-6.5)
{\includegraphics[scale = 0.82]{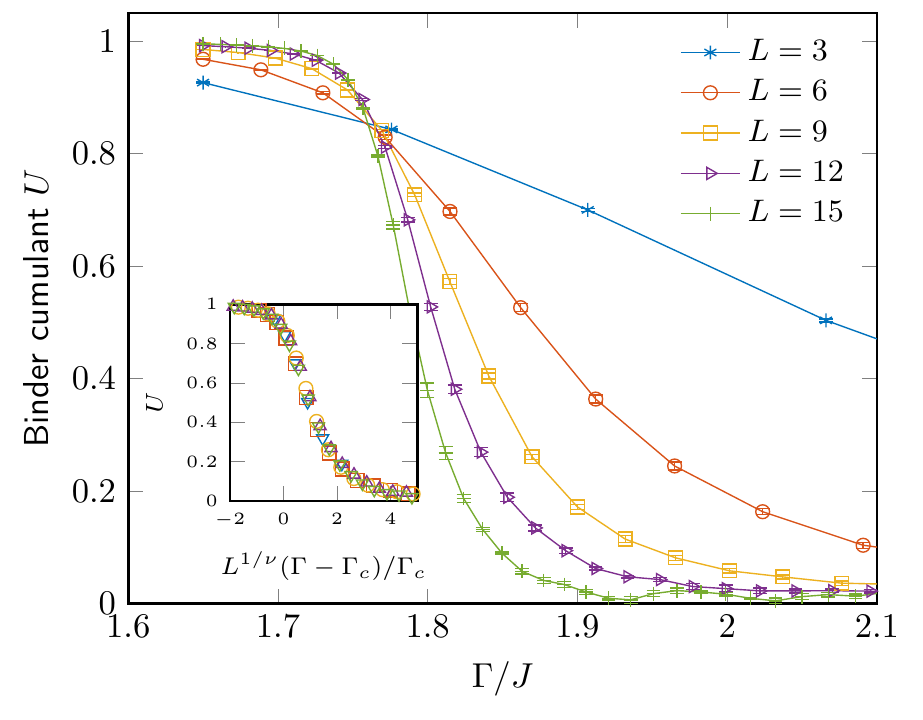}};
\node[anchor=north east] (phasediagram) at (11.5,-6.6)
{\includegraphics[scale=.82]{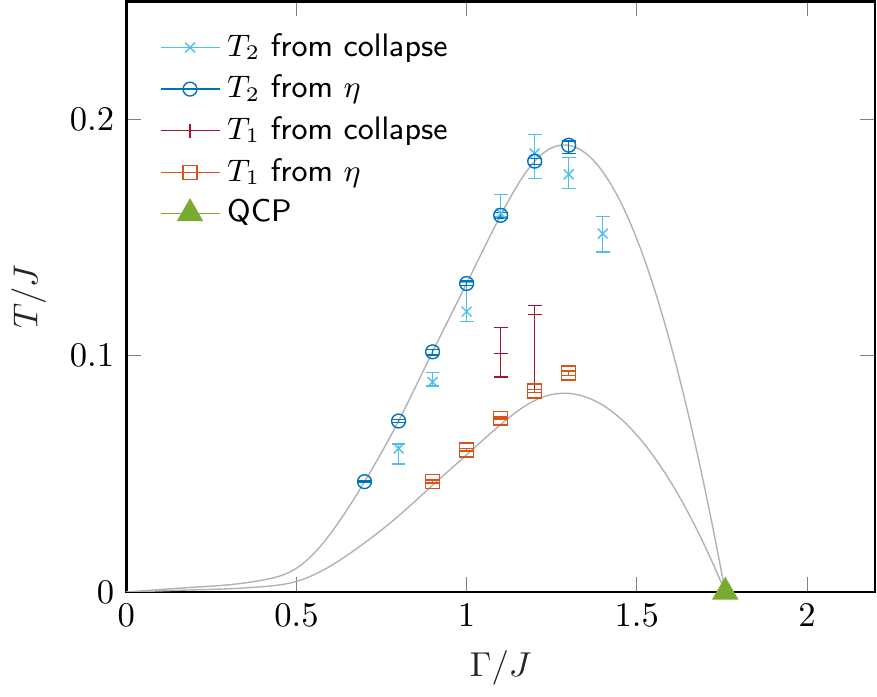}};
\node (a) at (-5,-.5) {{\bfseries\sffamily\large a}};
\node (b) at (1,-.5) {{\bfseries\sffamily\large b}};
\node (b) at (7,-.5) {{\bfseries\sffamily\large c}};
\node (b) at (-4.5,-7) {{\bfseries\sffamily\large d}};
\node (b) at (4.5,-7) {{\bfseries\sffamily\large e}};

\end{tikzpicture}\vspace{-.25cm}

\caption{{\bf Phase diagram of the square-octagonal lattice from Monte Carlo simulations with toroidal boundary conditions.}  {\bf a--b}, We estimate the upper ({\bf a}) and lower ({\bf b}) KT phase transition via universal collapse of susceptibility data, varying temperature for several values of $\Gamma/J$ (shown: $\Gamma/J=1.2$); lower collapse is imperfect.  {\bf c}, Power-law scaling of $m$ with $L$ at upper and lower critical temperatures with critical exponents from 2D XY universality.  Lines show power-law regression.  {\bf d}, Crossing of the normalized Binder cumulant for models with $\beta = 3.5 L/\Gamma$ gives an estimate of the quantum critical point $\Gamma/J \approx 1.76$ (inset: collapse of Binder cumulant across system sizes).  {\bf e}, From this data we derive a picture of the phase diagram of the square-octagonal lattice with $J_{\text{FM}} = -1.8J_{\text{AFM}}$.
 }\label{fig:phasediagram}

\end{figure}

First, we use the expected power-law decay of correlations---and consequently $m$---as system size increases.  Within the critical region and for fixed $\Gamma$, $m_L$ is expected to scale as $L^{-\eta/2}$, with critical exponents $\eta_1=1/9$ and $\eta_2=1/4$.  From this we determine $T_1$ and $T_2$ using a power-law fit on values of $L$ between $6$ and $21$ (Fig.~\ref{fig:phasediagram}).

Second, we additionally determine $T_1$ and $T_2$ by fitting our data to universal scaling curves.  For the lower transition, we have
\begin{equation}
  m_LL^{b} = m_0(L^{-1}e^{at^{-1/2}})
\end{equation}
where $m_0$ is a universal function, $a$ is a nonuniversal constant, and $t = (T_1-T)/T_1$ is the residual temperature approaching $T_1$ from below.  For the lower transition, we have
\begin{equation}
  \chi_LL^{-c} = \chi_0(L^{-1}e^{at^{-1/2}})
\end{equation}
where $\chi_0$ is a universal function, $a$ is a nonuniversal constant, and $t = (T-T_2)/T_2$ is the residual temperature approaching $T_2$ from above.  We expect from universality that in these fits, $b\approx (1/9)/2$ and $c \approx 7/4$.

For the upper transition, we obtain good collapses for $\Gamma$ between $0.7$ and $1.3$.  These values are in fairly good agreement with values of $T_1$ determined from $\eta$, as in the triangular lattice.\cite{Isakov2003}  For the lower transition, fitting is a generally poor and we only obtain a reasonably convincing collapse for $\Gamma \in [1.1,1.2]$ by discarding data for $L<15$.  Values of $T_1$ deviate significantly from those given by $\eta$.  A study of larger instances would clarify the picture.

\subsection{Quantum critical point}

Again extending from the triangular lattice, we anticipate a quantum phase transition consistent with the 3D XY universality class\cite{Blankschtein1984a,Isakov2003,Wang2017} with dynamical exponent $z=1$ and universal exponent $\nu = 2/3$.  To estimate the quantum critical point, we measure the normalized Binder cumulant
\begin{equation}
U = 2  \left(1 - \frac{\langle m^4 \rangle}{2\langle m^2\rangle^2}\right).
\end{equation}  We do so for system sizes up to $L=15$ using inverse temperature $\beta(L,\Gamma) = 3.5L^z/\Gamma$.  The Binder cumulant $U$ crosses near a critical point $\Gamma_c \approx 1.76$.  Scaling of $U$ in the vicinity of the quantum critical point collapses as $U_L = L^{1/\nu}(\Gamma-\Gamma_c)/\Gamma_c $ using $\nu = 2/3$.

\clearpage

\section{Effect of quench}\label{sec:quench}

\begin{figure*}
\begin{tikzpicture}\sf
\node[anchor=north west] (mversuss) at (0,0)
{\includegraphics[scale=.82]{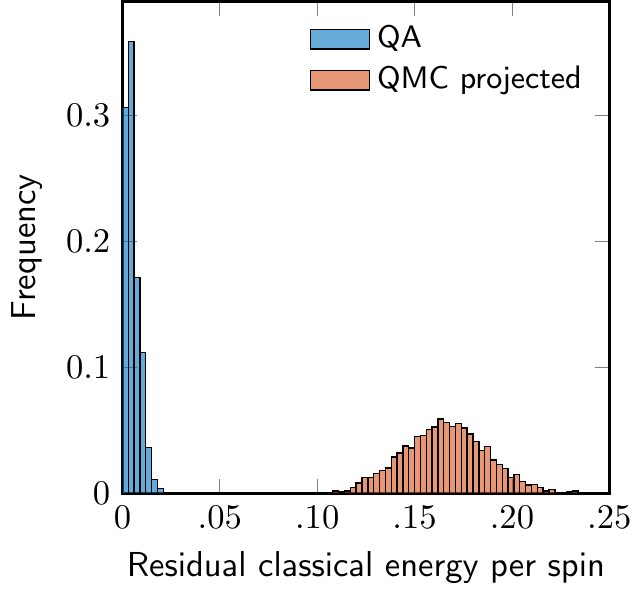}};
\node[anchor=north west] (mversuss) at (6,0)
{\includegraphics[scale=.82]{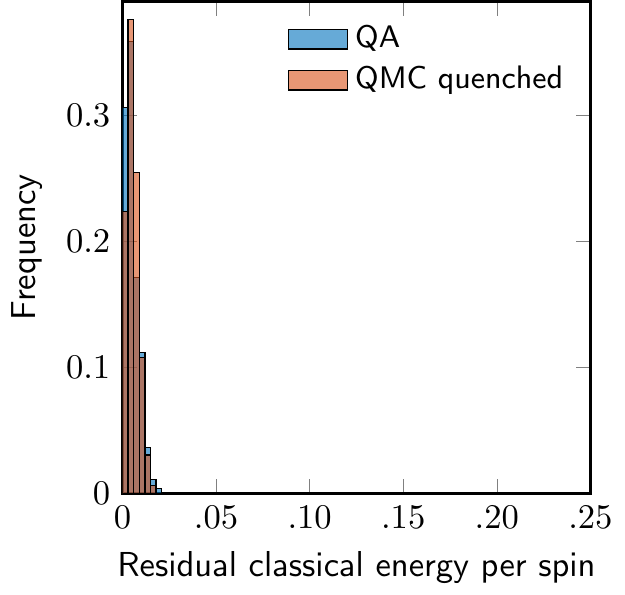}};
\node[anchor=north west] (mversuss) at (12,0)
{\includegraphics[scale=.82]{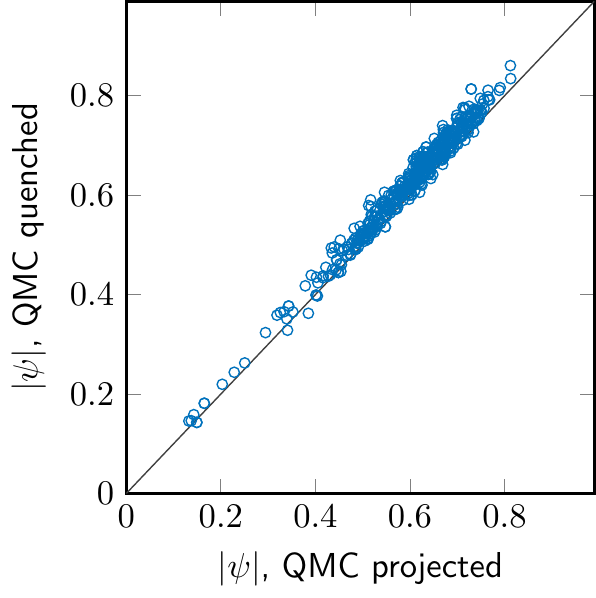}};
\node (a) at (0.5,-.1) {{\bfseries\sffamily\large a}};
\node (b) at (6.5,-.1) {{\bfseries\sffamily\large b}};
\node (b) at (12.5,-.1) {{\bfseries\sffamily\large c}};
\end{tikzpicture}\vspace{-.25cm}
\caption{{\bf Effect of classical quench on QMC samples.}  {\bf a}, Distributions of classical energy differ significantly between QA and QMC output, with QA giving much lower classical energies.  {\bf b}, When a local quench is performed on QMC output, removing local excitations at the single qubit level and the four-qubit chain level, the energy distribution matches QA closely.  {\bf c}, Quenching increases $\langle|\psi|\rangle$ by $0.02$ in these QMC samples.  Samples are collected for $s=0.26$, $T=\SI{8.4}{mK}$.}\label{fig:quenchenergy}
\end{figure*}

The QA simulation generally provides shows good agreement with QMC in estimates of $\langle m \rangle$, consistent with the formation and annihilation of vortex-antivortex pairs.  However, these excitations appear far less often in QA output states than in projected QMC states.  Here we show that this can be explained by evolution during the $\SI{1}{\micro\second}$ QA quench.  As quantum and thermal fluctuations are reduced, tightly-bound vortex-antivortex pairs are annihilated.  Figure \ref{fig:quenchenergy} compares the mean residual classical energy per spin of QA and QMC.  QMC projected states have many more excitations than QA states.  We model a local classical quench in QMC output by repairing classical excitations at the single-qubit level and four-qubit chain level.  After this classical quench is applied to QMC output, residual energies of QMC states resemble those of QA states.  Effect on $\psi$ is minimal.  This is consistent with the hypothesis that similar annihilation of defects occurs during the QA quench.  We therefore expect that faster QA quench should lead to greater population of classical excitations---vortices and antivortices---in QA output.

\end{document}